\documentclass[12pt]{article}

\usepackage{a4}
\usepackage{latexsym,amssymb,amsmath}
\usepackage{citesort}
\usepackage{psfrag}

\usepackage{graphicx}
\newcommand{\Span}{\text{\rm Span}}

\DeclareFontFamily{OT1}{rsfs}{} \DeclareFontShape{OT1}{rsfs}{m}{n}{
<-7> rsfs5 <7-10> rsfs7 <10-> rsfs10}{}
\DeclareMathAlphabet{\mycal}{OT1}{rsfs}{m}{n}
\def\scri{{\mycal I}}%
\def\scrip{\scri^{+}}%
\def\scrp{{\mycal I}^{+}}%
\def\Scri{\scri}

\newcommand{\ptc}[1]{\mnote{  #1}}

\newcommand{\cref}[1]{4\emph{\ref{#1})}}
\newcommand{\hS}{\hat{\Sigma}}
\newcommand{\hX}{\hat{X}}
\newcommand{\pS}{\partial\Sigma}
\newcommand{\ps}{\pS}
\newcommand{\dS}{\dot{\Sigma}}
\newcommand{\ds}{\dS}
\newcommand{\pdS}{\partial\dS}
\newcommand{\pds}{\pdS}
\newcommand{\pd}{\partial_{\doc}}
\newcommand{\ddoc}{\dot{\doc}}

\newcommand{\bcU}{\breve{\cal U}}
\newcommand{\cU}{{\cal U}}
\newcommand{\cN}{{\cal N}}
\newcommand{\cO}{{\cal O}}
\newcommand{\cNX}{{\cal N}_X}
\newcommand{\cS}{{\cal S}}
\newcommand{\cV}{{\cal V}}
\newcommand{\R}{\Bbb R}
\newcommand{\ext}{{\mathrm{ext}}}
\newcommand{\Sext}{\Sigma_\ext}
\newcommand{\Snd}{\partial_{nd}\Sigma}
\newcommand{\hg}{\hat \gamma}
\newcommand{\tg}{\tilde \gamma}
\newcommand{\tk}{\tilde \kappa}
\newcommand{\hI}{\hat I}

\newcommand{\communication}{{communications}}
\newcommand{\communications}{{communications}}
\newcommand{\proof}{\noindent{\sc Proof:}\ }

\newtheorem{Theorem}{Theorem}[section]
\newtheorem{Corollary}[Theorem]{Corollary}
\newtheorem{Lemma}[Theorem]{Lemma}
\newtheorem{Proposition}[Theorem]{Proposition}

\newtheorem{Remark}[Theorem]{Remark}

\newcommand{\Proof}{\proof}
\newcommand{\QED}
   {\hfill$\hbox{\vrule height1.3ex width1.3ex depth.1ex}\ $
     \\ \medskip}
\newcommand{\qed}{\QED}
\newcommand{\eq}[1]{(\ref{#1})}

 \def\scri{\hbox{${\cal J}$\kern -.645em {\raise
      .57ex\hbox{$\scriptscriptstyle (\ $}}}}

 \def\Scri{\scri}

\newcommand{\ks}{{\cal K}(\Sigma)}

\newcommand{\kds}{{\cal K}(\dS)}
\newcommand{\pkds}{\partial{\cal K}(\dS)}
\newcommand{\Mext}{M_{\mathrm{ext}}}
\newcommand{\doc}{{\cal D}_{oc}}
\newcommand{\pdoc}{\partial\doc}

\newcommand{\be}{\begin{equation}}
\newcommand{\ee}{\end{equation}}

{\catcode `\@=11 \global\let\AddToReset=\@addtoreset}
\AddToReset{equation}{section}
\renewcommand{\theequation}{\thesection.\arabic{equation}}

\newcounter{mnotecount}[section]

\renewcommand{\themnotecount}{\thesection.\arabic{mnotecount}}

\newcommand{\mnote}[1]
{\protect{\stepcounter{mnotecount}}$^{\mbox{\footnotesize  $
      \bullet$\themnotecount}}$ \marginpar{\raggedright\tiny\em
    $\!\!\!\!\!\!\,\bullet$\themnotecount: #1} }

\newcommand{\oldmnote}[1]{ \marginpar{\raggedright\tiny\em old mnote
  in the file here  (to be discarded as far as PC is   concerned)}}
\newcommand{\oldnote}[1]{}  
\newcounter{pcheckcount}[section]

\newcommand{\pcheck}[1]{}

\begin{document}

\title{The classification of static vacuum space--times containing an
  asymptotically flat spacelike hypersurface with compact interior}

\author{
Piotr T.\ Chru\'sciel\protect\thanks{%
  Alexander von Humboldt fellow.  Supported in part by a grant from
  the Polish Committee for Scientific Research (KBN) No 2 PO3B 073 15
  and by the Humboldt Foundation.
  \emph{Email}: Chrusciel@Univ-Tours.Fr} \\
D\'epartement de Math\'ematiques\\
Facult\'e des Sciences\\
Parc de Grandmont\\
F37200 Tours, France }
\maketitle

\begin{abstract}
  We prove non--existence of static, vacuum, appropriately regular,
  asymptotically flat black hole space--times with degenerate (not
  necessarily connected) components of the event horizon. This
  finishes the classification of static, vacuum, asymptotically flat
  domains of outer communication in an appropriate class of
  space--times, showing that the domains of outer communication of the
  Schwarzschild black holes exhaust the space of appropriately regular
  black hole exteriors.
\end{abstract}

\section{Introduction}
\label{SI}

 \ptc{The main text here should be essentially identical with
 the version published in~\cite{Chstatic}, and indeed with the
 original gr-qc v1 version, except for an extended
 bibliography, and  a few margin notes pointing out the
 existing problems, and an addendum at the end of the paper
 showing how those can be solved in the globally hyperbolic
 case. Furthermore, the numbering of references is different
 because of the new references from the Addendum.}
A classical question in general relativity, first raised and partially
answered by Israel \cite{Israel:vacuum}, is that of classification of
black hole solutions of the vacuum Einstein equations satisfying some
regularity conditions.  The most complete result existing in the
literature is that of Bunting and Masood--ul--Alam
\cite{bunting:masood} who show, roughly speaking, that all
appropriately regular such black holes which \emph{do not contain
  degenerate horizons} belong to the Schwarzschild family. In this
paper we remove the condition of non--degeneracy of the event horizon
and show the following: \vfill \eject
\begin{Theorem}
 \ptc{The proof as given in the main, original body of the paper does not correctly exclude the possibility
 that there could be components of $\partial \Sigma$ on which
 $\nabla (g(X,X))$ vanishes and on which $X$ has zeros;
 Proposition~\ref{Ponce2x} in the Addendum takes care of the
 problem.}
  \label{T1}
  Let $(M,g)$ be a static solution of the vacuum Einstein equations
  with defining Killing vector $X$. Suppose that $M$ contains a
  connected space-like hypersurface $\Sigma$ the closure $\bar \Sigma$
  of which is the union of a finite number of asymptotically flat ends
  and of a compact interior, such that:
\begin{enumerate}
 \item We have $g_{\mu\nu}X^\mu X^\nu < 0 $ on\footnote{We use the
    signature $(-,+,+,+)$.} $\Sigma$.
\item The topological boundary $\partial \Sigma\equiv
  \overline{\Sigma}\setminus \Sigma$ of $\Sigma$ is a nonempty
  topological manifold, with $g_{\mu\nu}X^\mu X^\nu = 0 $ on
  $\partial\Sigma$.
  \end{enumerate}
Then $\Sigma$ is diffeomorphic to $\R^3$ minus a ball, so that it is
simply connected, it has only one asymptotically flat end, and its
boundary $\partial \Sigma$ is connected. Further there exists a
neighborhood of $\Sigma$ in $M$ which is isometrically diffeomorphic
to an open subset  of the Schwarzschild space--time.
\end{Theorem}

The various notions used here are spelled out in detail in Section
\ref{SP} below.

Theorem \ref{T1} gives a complete classification of asymptotically
flat, static space--times with singularity--free space--like
hypersurfaces and with boundaries defined by the condition that the
Lorentzian norm squared of the Killing vector field vanishes there,
where the notion of ``singularity--free'' is made precise in the
statement above.  In fact, together with the Lichnerowicz
theorem\footnote{\label{lichne}We note that the sharpest version of
  the Lichnerowicz theorem currently available in the static case is
  that of \cite[Theorem 1.1]{manderson:static}.}  (reviewed in Section
\ref{Komar} below) it gives a complete classification of vacuum
space--times which contain an asymptotically flat spacelike hypersurface
$\Sigma$ with compact interior (the meaning of that notion should be clear
from the statement of Theorem \ref{T1} and from the discussion in Section
\ref{Komar} below) and a Killing vector field which is timelike on $\Sigma$
and satisfies the staticity condition. We note that we do \emph{not}
assume any causal regularity conditions on $(M,g)$ (in fact, even the
hypothesis of time orientability imposed in Section \ref{SP} below is not
needed here). The result is sharp: the extension of the Curzon space--time
constructed in \cite{ScottSzekeresII} contains space--like hypersurfaces
$\Sigma$ which satisfy all the hypotheses above except for the (implicit)
condition of compactness of $\partial \Sigma$. It has been conjectured by
M.~Anderson \cite[Conjecture 0.3]{manderson:static} that this
condition\footnote{I am grateful to M.~Anderson for useful
  comments concerning those points.} is not necessary when $\Sigma$ is
taken to be normal to the Killing vector field $X$.

A loose way of stating the main point of the above result, as compared
to the ones previously available, is that we are showing
non--existence of static, vacuum, regular black hole space--times with
degenerate (not necessarily connected) components of the event
horizon.  We further note that the Bunting and Masood--ul--Alam
version of the above result requires (in addition to the
non--degeneracy condition) $\Sigma$ to be normal to the Killing vector
field $X$. The hypothesis that $\Sigma$ is normal to the Killing
vector field has been relaxed under various further hypotheses,
including various global causality hypotheses on $(M,g)$
\cite{Sudarsky:wald,RaczWald2,ChWald1,Chnohair}, however no statement
with the generality above is available in the literature even in the
case where no degenerate horizons are present.

It might be of some interest to mention that our conclusion will still
hold for quite a larger class of manifolds $\Sigma$. A possible
generalization is that with $\Sigma$ being \emph{e.g.\/} the union of
a) a finite number of asymptotically flat ends with b) a neighborhood
of the boundary $\partial \Sigma$ which has compact closure and c) a
non--compact region on which we have $-1+\epsilon < g_{\mu\nu}X^\mu
X^\nu < -\epsilon $, provided that $\Sigma$ with the induced metric is
a \emph{complete} Riemannian manifold. The proof carries through
without any modifications to this case.

We note that the closure $\bar\Sigma$ in point 2 of Theorem \ref{T1}
(and everywhere else in this paper) is taken in the space-time $M$,
and that $\partial\Sigma$ defined in this way is sometimes called the
edge of $\Sigma$ in the physical literature. This should \emph{not} be
confused with the \emph{metric boundary} of $(\Sigma,\gamma)$, where
$\gamma$ is, \emph{e.g.}, the metric induced by $g$ on $\Sigma$, or
some other metric on $\Sigma$. By considering spacelike hypersurfaces
in Schwarzschild space--time it is easily seen that $\partial\Sigma$
will typically have corners at points at which the Killing vector
field vanishes, and is therefore \emph{not} a differentiable
submanifold of $M$ in such cases. Further, it is not \emph{a priori}
clear that $\Sigma$ can always be chosen so that $\partial \Sigma$ is
a smooth submanifold of $M$ and/or of $\bar \Sigma$ even at points at
which $X$ does not vanish.

Our strategy is essentially the same as that of Bunting and
Masood--ul--Alam \cite{bunting:masood}, though our starting points
differ: while Bunting and Masood--ul--Alam consider the metric induced
on the hypersurface normal to the Killing vector field, we consider
the \emph{orbit space} metric $h$ on $\Sigma$, as defined in Section
\ref{SBc} below.  The key first step, which is new, is the analysis of
the geometry of $(\Sigma,h)$ near both the degenerate components of
$\partial\Sigma$ (\emph{cf.\/} Proposition \ref{pbc1}) and the
non--degenerate ones (\emph{cf.\/} Proposition \ref{pbc2}). Next,
following \cite{bunting:masood}, we consider a manifold which consists
of two copies of $(\Sigma,h)$ glued along all \emph{non--degenerate}
components of $\partial \Sigma$, equipped with an appropriate
conformally deformed metric. The key next element of our proof is a
new version of the positive energy theorem proved
in\footnote{\label{geroch}Once this paper was written we have realized
  that Theorem \ref{tpet} can also be inferred from \cite[Theorem
  6]{GerochPerng}.} \cite{BartnikChrusciel1} (\emph{cf.\/} Theorem
\ref{tpet} below). Using those results one shows that the metric on
$\Sigma$ is conformally flat. One can then use classical arguments to
finish the proof\footnote{We note that it is usual in this last step
  of the proof to invoke analyticity to conclude. Because analytic
  extensions of manifolds are not unique this is not sufficient
  without a more thorough justification.} (\emph{cf., e.g.},
\cite[Section II]{Lindblom}, together with \cite[Section 3]{Kunzle} or
\cite[Lemma 4]{BeigSimon3}); we present here a new argument,
essentially due to Herzlich\footnote{We are grateful to M.~Herzlich
  for allowing us to reproduce his unpublished proof here.}
(M.~Herzlich, private communication), which gives a considerably
simpler proof of this last step and avoids the problems related to
uniqueness of analytic extensions.

Under the hypotheses of Theorem \ref{T1} there is no chance of getting
more information about the size of the set on which the metric is that
of a Schwarzschild space--time (consider any hypersurface $\Sigma$ in
the Schwarzschild--Kruszkal--Szekeres space--time, and set $M$ to be
any neighborhood of $\Sigma$ which does not coincide with the
Schwarzschild space--time; alternatively, identify $t$ with $t+1$ in
the Schwarzschild space--time). Thus, to get more information about
the size of this set some more hypotheses are needed. A simplest
result of this kind is the following:

\begin{Corollary}\label{c1}
  Under the hypotheses of Theorem \ref{T1}, assume further that
  \begin{enumerate}
  \item[3.]  The orbits of the Killing vector $X$ through $\Sigma$ are
    complete.
  \end{enumerate}
 Then
  the following properties are equivalent:
  \begin{enumerate}
  \item[\phantom{ii}i.] $\Sext$ is achronal$ \!$ {}\footnote{By that
      we mean that there are no timelike curves from $\Sext$ to itself
      which are entirely contained in $\Mext$.} in $\Mext$.
  \item[\phantom{i}ii.] $\Mext$ is diffeomorphic to $\R\times\Sext$ (which is
    equivalent to $\scri$ having $\R\times S^2$ topology).
  \item[iii.] There are no closed timelike curves through $\Sext$
    contained in $\Mext$.
  \end{enumerate}
  Further, if one (and hence all) of the above conditions holds, then
  the Killing development$\!$ \footnote{The notion of Killing
    development used here differs slightly from the definition of
    \cite{ChBeig1} as we allow here a topology of $\ks$ which is not
    $\R\times\Sigma$.} $\ks$ of $\Sigma$ defined as
\begin{equation}
\label{kddef}
\ks\equiv\cup_{t\in \R} \phi_t(\Sigma)\ ,
\end{equation}
where $\phi_t$ is the action of the isometry group generated by $X$,
equipped with the induced metric, is isometrically diffeomorphic to a
domain of outer communications in the
Schwarzschild--Kruszkal--Szekeres space--time.
\end{Corollary}

The definition of the domain of outer communications used here is
given in Section \ref{SP} below.

Strictly speaking, both Theorem \ref{T1} and Corollary \ref{c1} are
not statements about black holes, because it is not \emph{a priori}
clear what is the relationship between their hypotheses and the
existence of a black hole region.  Moreover, in Corollary \ref{c1} it
is not clear how much of the space--time is covered by the Killing
development of $\Sigma$. Now, there are various goals one might wish
to achieve: a) one might rest content with Corollary \ref{c1} (this is
indeed suggested by the relatively weak hypotheses thereof); b) one
might want to show that the d.o.c. of $(M,g)$ is isometrically
diffeomorphic to a d.o.c. of the Schwarzschild--Kruszkal--Szekeres
space--time; c) one might wish to show that $(M,g)$ is the
Schwarzschild--Kruszkal--Szekeres space--time.

Concerning b) above, we are not aware of any construction which would
show that more hypotheses than those of Corollary \ref{c1} are needed
to obtain this conclusion. Let us however note that there is such an
example in electro--vacuum space--times, which is obtained as follows:
let $(\hat M,\hat g)$ be the extension of the Reissner--Nordstr\"om
space--time described by the Carter--Penrose diagram on p.~158 of
\cite{HE}. The diffeomorphism $\psi$ obtained by mapping a point 
on this diagram to a point ``shifted by two blocks up'' is an isometry
of this space. Then $(M=\hat M/\psi, g)$, where $g$ is the obvious
metric on $M$, is an electro--vacuum space--time in which we can find
a hypersurface $\Sigma$ satisfying the conditions of Corollary
\ref{T1}. (We note that in this space--time there are closed
time--like curves through every point, but there are \emph{no closed
  timelike curves through $\Sext$ contained in $\Mext$}.) The d.o.c.
associated with any asymptotically flat region $\Mext$ in $M$ is the
whole space--time $M$, and is therefore \emph{not} isometrically
diffeomorphic to a d.o.c. in $(\hat M,\hat g)$ (which consists of only
one of the blocks of the Carter--Penrose diagram on p.~158 of
\cite{HE}).

While this example does not satisfy the vacuum Einstein equations, it
clearly shows that the standard global techniques of Lorentzian
geometry, which assume at most some energy inequalities, cannot be
sufficient to achieve the conclusion that the d.o.c. of $(M,g)$ is
isometrically diffeomorphic to a d.o.c.  of the
Schwarzschild--Kruszkal--Szekeres space--time. Thus some further
hypotheses need to be imposed unless a careful study of the vacuum
field equations is performed. We note that invoking analyticity will
not help, unless one has a proof that the metric has to be analytic up
to and beyond the event horizon. In any case non--uniqueness of
analytic extensions of Lorentzian manifolds leads to problems even if
one assumes that the whole space--time is analytic.  For example, the
vacuum examples constructed in \cite{Ch:rigidity} show that neither
analyticity alone, nor
analyticity together with the set of hypotheses in \cite{HE} will 
suffice to conclude that $(M,g)$ must be the
Schwarzschild--Kruszkal--Szekeres space--time. We further note that
it is certainly of interest to completely classify the exterior
regions of black hole space--times, while it is perhaps of limited
interest to try to set up some heavy set of hypotheses which will
allow us to say something about what is happening beyond the event
horizons. In any case b) and c) are separate issues.  Here we will
ignore question c) and only address question b), using the usual
global Lorentzian techniques, and prove the following result:

\begin{Theorem}
  \label{T1.1}
  Let $(M,g)$ be a  solution of the vacuum Einstein
  equations containing a connected space-like hypersurface $\Sigma$,
  the closure $\bar \Sigma$ of which is the union of a finite number
  of asymptotically flat ends and of a compact interior.  Let $X$ be a
  Killing vector field on $M$ which is timelike, future directed in
  all the asymptotically flat ends and satisfies the staticity
  condition \eq{eq:st}.  Let further $\doc\equiv \doc(\Mext)$ be a
  domain of outer communications in $(M,g)$ associated to one of the
  asymptotically flat ends of $\Sigma$. Suppose that:
\begin{enumerate}
\item We have $\Sigma\subset \doc$.
\item The topological boundary $\partial \Sigma\equiv
  \overline{\Sigma}\setminus \Sigma$ of $\Sigma$ is a nonempty
  topological manifold and satisfies $\partial \Sigma = \overline
  \Sigma \cap \pdoc$.
\item
 $X$ has  complete orbits in $\doc$. 
\end{enumerate}
\renewcommand{\theenumi}{\alph{enumi}}
\renewcommand{\labelenumi}{4\theenumi)} In addition to the above,
suppose that one of the following conditions holds:
\begin{enumerate}
\item \label{first} 
\label{docgh} Either   $(\doc,g|_{\doc})$ is globally hyperbolic, or
\item \label{Mgh} $(M,g)$ is globally hyperbolic, or

  \ptc{A correct proof of points
  4\ref{simply}-4\ref{nozeros} requires the supplementary
  hypothesis of non-existence of degenerate non-embedded
  prehorizons within $\doc$. Under the hypotheses of points
  4\ref{docgh} or 4\ref{Mgh} this property is established
 in~\cite{ChGstaticity}, and holds trivially under the
 hypotheses of point 4\ref{last}. Note that a supplementary
 hypothesis of analyticity of the metric on $\doc$ would
 immediately imply non-existence of such prehorizons.}
\item \label{simply} there are no closed timelike curves through
  $\Sext$ contained in $\doc$,  with $\doc$ being moreover simply
  connected, or
\item \label{simplynew} $\doc\setminus\{X=0\}$ is simply connected, or
\item \label{nowhite} $\Sigma$ is achronal$ \!\!$ {}\footnote{By that
    we mean that there are no timelike curves from $\Sigma$ to itself
    which are entirely contained in $\doc$.} in $\doc$ and the white
  hole region (\emph{cf.\/} Equation \eq{wh}) is empty, or
\item \label{nozeros} $\Sigma$ is  achronal in $\doc$ and $X$
  has no zeros on $\partial\Sigma$, or
\item \label{allzeros} $\Sigma$ is  achronal in $\doc$ and $X$
  is nowhere light-like on $\overline\Sigma$.
\label{last} 
\end{enumerate}
Then the conclusions of Theorem \ref{T1} and Corollary \ref{c1} hold.
Moreover $\doc$ is isometrically diffeomorphic to a domain of outer
communications of the Schwarzschild--Kruszkal--Szekeres space--time.
\end{Theorem}

The set of hypotheses \cref{first}--\cref{last} will be referred to as
\emph{alternative} hypotheses, while the remaining conditions will be
referred to as the \emph{main} hypotheses.

To avoid ambiguities, we emphasize that in Theorem \ref{T1.1} it is
\emph{not} assumed that $X$ is timelike throughout $\Sigma$.

It should be clear from the long list of alternative conditions we
have given that we are not satisfied with any single one of them. We
note that the alternative hypothesis \cref{docgh} is rather natural in
many global problems in general relativity. It has the elegant feature
that it makes no hypotheses on the global causal structure of $M$ away
from $\doc$. It is necessary since any d.o.c. of the
Schwarzschild--Kruszkal--Szekeres space--time has this property. (It
can be pointed out that each of the alternative conditions above
eventually implies that $\doc$ is globally hyperbolic, hence that the
alternative condition \cref{docgh} holds.) This is the hypothesis that
we consider to be the most satisfactory amongst all alternative
hypotheses above.  Nevertheless, in view of the rather weak conditions
of Corollary \ref{c1} one is tempted to look for \emph{a priori}
weaker conditions which would lead to a theorem of the kind of Theorem
\ref{T1.1}.  We conjecture that the hypothesis that $\Sigma$ is
achronal in $\doc$ should be enough
for the conclusion of Theorem \ref{T1.1} to hold, thus the remaining
restrictions in \cref{nowhite}--\cref{allzeros} are unnecessary.

 We further note the following:

\begin{enumerate}
\item While the alternative hypothesis \cref{Mgh} might also be
  natural for several purposes, we note that it seems to exclude
  degenerate horizons at the outset. For example, consider the
  extension $(\hat M,\hat g)$ of the extreme Reissner--Nordstr\"om
  space--time as described by the Carter--Penrose diagram given on p.
  160 of \cite{HE}. By inspection of this diagram one easily finds
  that the only globally hyperbolic subset $(M,g)$ of $(\hat M,\hat
  g)$ which contains asymptotically flat hypersurfaces and has
  complete Killing orbits has to be one of the d.o.c.'s of that
  space--time, hence will have no horizons, no black--hole regions,
  and no hypersurfaces as required in the main hypotheses of Theorem
  \ref{T1.1}. (We note, however, that a space--time with the global
  structure of $(\hat M,\hat g)$ could \emph{a priori} satisfy the
  main hypotheses above together with any alternative hypotheses other
  than \cref{Mgh} and \cref{allzeros}.)
\item The alternative hypothesis \cref{simply} together with the
  accompanying proof is (up to minor improvements) due to Carter
  \cite[Theorem 4.1]{CarterlesHouches}. We do not find the hypothesis
  \cref{simplynew} especially natural, we have added it for
  completeness as it is closely related to the hypothesis
  \cref{simply} while being shorter to formulate.
\item Theorem \ref{T1.1} with the alternative hypothesis
  \cref{nowhite} gives a classification of appropriate space--times
  which have no white hole regions -- that is, ``pure black hole''
  solutions.
\item The requirement that $\Sigma$ be normal to the Killing vector
  field, which occurs in the original form of the
  Israel--Bunting--Masood--ul--Alam theorem, implies that the
  alternative hypothesis \cref{allzeros} holds; as already mentioned,
  it excludes degenerate horizons at the outset.
 \end{enumerate}

 We note that while Theorems \ref{T1} and \ref{T1.1} are both new in
 the generality given here, the transition from Theorem \ref{T1} to
 Theorem \ref{T1.1} is in principle known. We shall give complete
 proofs because the arguments needed are scattered across the
 literature
 \cite{HE,CarterlesHouches,Wald:book,ChWald1,ChAscona,ChNad}, and
 because those arguments are often carried out under hypotheses which
 are different from the ones made here.

This paper is organized as follows: Section \ref{SP} contains
definitions and some preliminary remarks. In Section \ref{SBc} we
analyze the boundary conditions satisfied by the orbit--space metric
near Killing horizons. In that section neither staticity nor energy
inequalities are assumed to hold. In Section \ref{Komar} we recall an
elementary and well known proof of Theorem \ref{T1}, based on the
Komar identity, under the hypothesis that all horizons are degenerate.
Actually in that proof asymptotically flat stationary space--times are
allowed provided that all the Killing horizons are
\emph{non--rotating}, and some further conditions are satisfied; this
is discussed in detail there. Theorem \ref{T1} is proved in Section
\ref{ptt1}, while Theorem \ref{T1.1} is proved in Section
\ref{ptt1.1}. We close the paper with some concluding remarks in
Section \ref{conclusions}.

\medskip

\noindent\textbf{Acknowledgments:} The author wishes to
emphasize the key contribution of R.~Bartnik to the results
presented here through the joint proof of the version of the
positive energy theorem in \cite{BartnikChrusciel1}. He
acknowledges useful discussions with or comments from
M.~Anderson, R.~Bartnik, R.~Beig, H.~Friedrich, G.~Galloway,
W.~Simon, R.~Wald and G.~Weinstein.
 \ptc{Furthermore, I am grateful to J.L.~Costa for pointing out
 the problem with Lemma~\ref{lcvnew}, addressed in the
 Addendum.}

\section{Preliminaries}
\label{SP}
All the manifolds 
are assumed to be paracompact, Hausdorff and smooth. Space--times are
equipped with smooth metrics
and are always assumed to be 
time--orientable.

The Hawking--Ellis \cite{HE} notation for causal futures
$J^{\pm}(\Omega)$, chronal futures $I^{\pm}(\Omega)$, \emph{etc.}, is
used throughout.  Further, whenever needed, we use the notation
$I^+(A;\Omega)$ to denote the chronological future of a set $A$ in a
space--time $\Omega$, \emph{etc.}


A space--like hypersurface $\Sigma _{\mathrm{ext}}$ will be called an
\emph{asymptotically flat end} if it is diffeomorphic to ${\Bbb R}^3$
minus a ball and if the fields $(g_{ij},K_{ij})$ induced on $\Sigma
_{\mathrm{ext}}$ by the space--time metric satisfy the fall--off
conditions
\begin{equation}
|g_{ij}-\delta _{ij}|+r|\partial _\ell g_{ij}|+\cdots +r^k|\partial _{\ell
_1\cdots \ell _k}g_{ij}|+r|K_{ij}|+\cdots +r^k|\partial _{\ell _1\cdots \ell
_{k-1}}K_{ij}|\le C_{k,\alpha}r^{-\alpha }\ ,  \label{falloff}
\end{equation}
for some constants $C_{k,\alpha}$, $\alpha >0$, $k\ge 1$. We shall
always implicitly assume $\alpha >1/2$ when the ADM mass will be
invoked, as this condition makes it well defined in vacuum.  It
follows in any case from \cite{KennefickMurchadha} or from
\cite[Section 1.3]{Chnohair} that in stationary vacuum space--times
there is no loss of generality in assuming $\alpha= 1$, $k$ --
arbitrary. A hypersurface will be said to be \emph{asymptotically
  flat} if it contains an asymptotically flat end $\Sigma
_{\mathrm{ext}}$.

A space--time $(M,g)$ containing an asymptotically flat end $\Sext$
will be called \emph{static} if there exists on $M$ a Killing vector
field $X$ which
is \emph{timelike} in  $\Sext$
and which satisfies
\begin{equation}
  \label{eq:st}
  X_{[\alpha}\nabla_\beta X_{\gamma]} = 0 \ .
\end{equation}
It can be shown under fairly weak hypotheses \cite{ChBeig1,ChBeig2}
(which are satisfied under the hypotheses of Theorem \ref{T1} by
\cite{Herzlich:mass}) that $\Sext$ can be boosted so that $X$
asymptotes the unit future directed normal to $\Sext$, and we shall
always assume that this is the case. A Killing vector field satisfying
the above requirements will be called the \emph{defining} Killing
vector field of the static space--time.  We emphasize that unless
explicitly indicated otherwise we do \emph{not} assume that the orbits
of $X$ are complete in $M$.

Let $X$ be a Killing vector field which asymptotically approaches the
unit normal to $\Sigma _{\mathrm{ext}}$ in an asymptotically flat end
$\Sigma _{\mathrm{ext}}$. Passing to a subset of $\Sigma
_{\mathrm{ext}}$ we can without loss of generality assume that $X$ is
time-like on $ \Sigma _{\mathrm{ext}}$, and we shall always assume
that this is the case.  If the orbits of $X$ through $ \Sigma
_{\mathrm{ext}}$ are complete, then an exterior four--dimensional
asymptotically flat region can be obtained by moving $\Sigma
_{\mathrm{ext}}$ around with the flow $\phi _t$;
\begin{equation}
M_{\mathrm{ext}}=\cup _{t\in {\Bbb R}}\phi _t(\Sigma _{\mathrm{ext}})\
.  \label{mext}
\end{equation}
Following \cite{ChWald1}, the {\em domain of outer \communications\/}
(d.o.c.) $\doc(\Mext)$ associated with $\Sigma _{\mathrm{ext}}$ or
with $M_{\mathrm{ext}}$ is then defined as
\begin{equation}
 \doc(\Mext)=J^{+}(M_{\mathrm{ext}})\cap J^{-}(M_{\mathrm{ext}})=
I^{+}(M_{\mathrm{ext}})\cap I^{-}(M_{\mathrm{ext}})
\ .
\label{doc}
\end{equation}
(The equality $J^{\pm}(M_{\mathrm{ext}})=I^{\pm}(M_{\mathrm{ext}})$ is
easily verified; \emph{cf., e.g.}, \cite[Section 12.2]{Wald:book} for a proof
in a \Scri\ context.)  It is shown\footnote{We note that
  Conjecture 1.8 of \cite[Section 1.3]{ChAscona}, needed for this
  equivalence, has been settled in \cite{ChBeig1}.}, under appropriate
conditions, in \cite[Section 1.3]{ChAscona} that for stationary vacuum
space--times the above definition of the domain of outer \communication\
is equivalent to the standard one using $ \hbox{${\cal
    J}$\kern -.645em {\raise .57ex\hbox{$\scriptscriptstyle (\ $}}}$
({\em cf.\ e.g.\/} \cite {HE,Wald:book}).  The definition \eq{doc} turns out to
be more convenient for many purposes.

The black hole region ${\cal B}$ associated with the asymptotic end $
\Sigma _{{\ext}}$ or with $M_{{\ext}}$ is defined as
\begin{equation}
{\cal B}=M\setminus J^{-}(M_{\mathrm{ext}})=M\setminus
I^{-}(M_{\mathrm{ext}}) \ ,  \label{bh}
\end{equation}
while the white hole region ${\cal W}$ associated with the asymptotic end $
\Sigma _{{\ext}}$ or with $M_{{\ext}}$ is defined as
\begin{equation}
{\cal W}=M\setminus J^{+}(M_{\mathrm{ext}})=M\setminus
I^{+}(M_{\mathrm{ext}})\ .  \label{wh}
\end{equation}
 Thus the occurrence of boundaries of
$\doc(\Mext)$ signals that of black hole or white hole regions.

To avoid ambiguities, we define the \emph{Schwarzschild} space--time
$(M^{\mathrm{Schw}},g^{\mathrm{Schw}})$ to be the manifold
$\{t\in\R,r\in (2m,\infty), q\in S^2\}$, with the metric
\begin{equation}
  \label{eq:schw}
  g^{\mathrm{Schw}}=- (1-\frac{2m}{r})dt^2 +
(1-\frac{2m}{r})^{-1}dr^2+r^2d\Omega^2\ ,
\end{equation}
where $d\Omega^2$ is the standard round metric on a unit
two--dimensional sphere $S^2$. We will refer to those coordinates as
the standard coordinates on the Schwarzschild space--time. We shall
call a \emph{Schwarzschild--Kruszkal--Szekeres space--time} the
extension of $(M^{\mathrm{Schw}},g^{\mathrm{Schw}})$ described
\emph{e.g.\/} by the Carter--Penrose diagram on page 154 of \cite{HE}.
We note that each of the two copies of
$(M^{\mathrm{Schw}},g^{\mathrm{Schw}})$ which can be seen on that
diagram forms a d.o.c with respect to the appropriate asymptotic
region. In Section \ref{ptt1} we shall need the so--called
\emph{isotropic} coordinates on the Schwarzschild space--time $(t,\bar
r,q)\in \R\times (m/2,\infty)\times S^2$, with $\bar r$ defined via
the equation $r=\bar r (1+m/(2\bar r))^2$, in which the Schwarzschild
metric takes the form
\begin{equation}
  \label{eq:confs}
  g^{\mathrm{Schw}}=- \Big(\frac{1-m/2\bar r}{1+m/2\bar r}\Big)^2dt^2 +
\Big(1+\frac{m}{2 \bar r}\Big)^{4}\Big(d\bar r^2+\bar r^2d\Omega^2\Big)\ .
\end{equation}

\section{Boundary conditions at Killing horizons}
\label{SBc}

In this section we shall consider Killing horizons in
\emph{stationary}\footnote{Part of the results presented in this
  section have been originally obtained under the hypothesis that $X$
  satisfies the staticity condition. G.~Weinstein pointed out to us
  that this analysis carries over to the stationary case when $h$ is
  interpreted as the orbit space metric, as defined below. This remark
  provided the breakthrough which led to the proof of the Lemma
  \ref{pbc3} below. We are grateful to him for this suggestion, and
  for several useful discussions.}  space--times; thus we shall
\emph{not} assume that the staticity condition \eq{eq:st} holds.
Further \emph{no field equations} or \emph{energy inequalities} are
assumed in this section.  A \emph{null hypersurface} $\cN$ will be
called a \emph{Killing horizon} if
\begin{enumerate}
\item $X$ is nowhere vanishing on $\cN$, and
\item $g_{\alpha\beta}X^\alpha X^\beta\equiv0$ on  $\cN$.
\end{enumerate}
In particular $X$ is necessarily tangent to the generators of $\cN$.
We shall sometimes write $\cNX$ for $\cN$ to emphasize that the
Killing horizon in question is associated with the given Killing
vector field $X$.  (The reader should be warned that what is usually
called a ``bifurcate horizon'' (\emph{cf., e.g.}, \cite{KayWald}) is
\emph{not} a Killing horizon in our terminology, rather it is the
union of four Killing horizons and of the ``bifurcation surface''.)
Recall that the surface gravity $\kappa$ of a Killing horizon is
defined by the formula
\begin{equation}
  \label{kdef}
  (X^\alpha X_\alpha)_{,\mu}\Big|_{\cNX} = -2\kappa X_\mu \ .
\end{equation}
A Killing horizon  $\cNX$ is said to be degenerate if $\kappa$ vanishes
throughout  $\cNX$.
From what is said in \cite{Boyer} one can infer the following:

\begin{Theorem}[Boyer \cite{Boyer}] \label{tboyer} Let  $\cN$ be a
  $C^1$ Killing horizon, with $X$ tangent to the generators of ${\cal
    N}$. Then:
  \begin{enumerate}
  \item $X$ is nowhere vanishing on the closure
      $\overline{\cal N}$ of any degenerate connected
      component of $\cal N$. \ptc{This is wrong, a
      counterexample is given in the Addendum. The error
      does not lie with \cite{Boyer}, but with the author
      of the current work. Proposition~\ref{Ponce2x} of the
      Addendum provides a correct statement needed for the
      remaining arguments in this work}
\item Let $p\in \overline{\cN}$ satisfy $X(p)=0$, set
  \begin{equation}
    \label{eq:bifdef}
    \cS=\{p\, |\, X(p)=0 \}\ .
  \end{equation}
  Then there exists a neighborhood $\cV$ of $p$ such that the set
  $\cS\cap \cV$ is a smooth, embedded, space--like, two--dimensional
  submanifold of $M$. Moreover the null geodesics normal to $\cS\cap
  \cV$ are Killing orbits such that $\cS\cap \cV$ is the accumulation
  set of those orbits in $\cV$.
  \end{enumerate}
\end{Theorem}

The set $\cS\cap \cV$ of point 2 above is usually called a
\emph{bifurcation surface} of the Killing horizon $\cN$. A good model
for this behavior is provided by the set of zeros of the usual
Killing vector field $\partial/\partial t$ in the
Schwarzschild--Kruszkal--Szekeres space--time.

In this section we wish to analyze the behavior near a Killing horizon
$\cN$ of the {orbit space metric} on a space--like hypersurface
$\Sigma$ such that $g_{\mu\nu}X^\mu X^\nu < 0 $ on $\Sigma$, with
$\partial \Sigma$ being a subset of the closure $\overline\cN$ of
$\cN$.  Let us start by defining the orbit space metric. Consider a
point $p\in M$ such that $g_{\mu\nu}X^\mu X^\nu(p)\ne 0$, then we have
the decomposition
$$T_pM = L(X)\oplus X^{\perp},$$ where $L(X)$ is the vector space
spanned by $X(p)$ and $X^{\perp}$ the space orthogonal to $X$. Let $Y$
be a vector tangent to $M$ at $p$, we can thus write
$$Y=Y_{\parallel}+ Y_\perp\ ,
$$
with self--explanatory notation. We define the \emph{orbit space
  metric} $h$ at $p$ by the formula
$$ h(Y,Z)=g(Y_\perp,Z_\perp)\ .$$ (We note that $h$ coincides at $p$
with the metric induced by $g$ on any hypersurface $N$ which has the
property that $T_pN= X^\perp$; this fact will play some role later.
However, we do \emph{not} assume that $\Sigma$ has this property.
Similarly in this section we do \emph{not} assume that $X^\perp$ forms
an integrable distribution. Finally we stress that the metric $h$
should \emph{not} be identified with the (natural) \emph{metric on the
  space of orbits} because we are \emph{not} assuming any regularity
properties of that space; in particular we do \emph{not} assume that
the space of orbits is a differentiable manifold, which is a minimum
requirement for the introduction of a metric on it.)  We have
$$ Y_\perp = Y - \frac{g(X,Y)}{g(X,X)} X\ , $$
so that
\begin{equation}
h(Y,Z)=g(Y,Z)- \frac{g(X,Y)g(X,Z)}{g(X,X)}\ .
 \label{eq:hdefnew}
\end{equation}
Consider, first, the case in which the connected component $S$ of
$\partial \Sigma$ under consideration corresponds to a degenerate
component of the event horizon:
\begin{equation}
  \label{eq:deg}
  \kappa\Big|_S = 0\ .
\end{equation}
We have the following:
\begin{Proposition}
  \label{pbc1} \ptc{the proof given here requires absence of
  zeros of $X$ on $\partial \Sigma$, which has been
  ``justified" by the incorrect point 2 of
  Theorem~\ref{tboyer}. An example of a degenerate Killing
  horizon with zeros of $X$ is given in the Addendum, but in
  that example $X$ is spacelike on both sides of $\partial
  \Sigma$, so this does not provide a counterexample to the
  statement, which we find likely to be true.
  Proposition~\ref{Ponce2x} in the Addendum proves
  non-existence of zeros of $X$ on $\partial\Sigma$ under the
  supplementary condition of hypersurface-orthogonality of $X$
  (which is sufficient for our purposes in this work), or of
  smoothness of $\partial\Sigma$.}
   Let $\Sigma$ be a $C^3$ space--like hypersurface in a
  space--time $(M,g)$ with Killing vector $X$, suppose that $\Sigma$
  is $C^3$ and space--like up to\footnote{\label{upto}Throughout this
    work ``up to'' means ``up to and including''.} its boundary $\pS$,
  with
  $$g_{\mu\nu}X^\mu X^\nu < 0 \ \mathrm{on}\ \Sigma, \quad
  g_{\mu\nu}X^\mu X^\nu = 0 \ \mathrm{on}\ \partial \Sigma\ .$$
  Then
  every compact connected component $S$ of $\pS$ which intersects a
  $C^2$ degenerate Killing horizon $\cNX$ corresponds to a
  \emph{complete} asymptotic end of $(\Sigma,h)$.
\end{Proposition}

{\noindent{\bf Remark:}\ } We note a discrepancy between the
degree of differentiability of $\cNX$ assumed here and that
asserted in Proposition \ref{Pkilling} below. It is conceivable
that with some effort one could weaken the hypotheses of
Proposition \ref{pbc1} --- we have not attempted to do that, as
this is irrelevant for the purpose of this paper. In view of
potential applications of Proposition \ref{pbc1} to the
classification of stationary black holes it might be of
interest to fill this gap.

\medskip

\proof Let $\cNX^0$ be the connected component of $\cNX$ which
intersects $S$. Connectedness of $S$ and of $\cNX^0$ shows that we
must have $\partial\Sigma \subset \overline{\cNX^0}$.  Boyer's Theorem
\ref{tboyer} implies that the Killing vector field $X$ is nowhere
vanishing on $S$, which leads to $\partial\Sigma \subset {\cNX}$.

Consider any one--sided neighborhood ${\cal O}\subset
\overline{\Sigma}$ of $S$ covered by coordinates $(y^i)=(x,v^A)$ in
which $S$ is given by the equation $x=0$; passing to a subset of
${\cal O}$ if necessary we can without loss of generality assume that
${\cal O}$ has compact closure. Let $\phi_s$ denote the (perhaps
locally defined) flow of the Killing vector field $X$ and set
$$
{\cal U} = \cup_{s\in (-\epsilon,\epsilon)} \phi_s({\cal O}) \ .
$$
Now the Killing vector field $X$ is non--spacelike
in ${\cal U}$, hence transverse to ${\cal O}$, so that the flow
parameter $s$ along the orbits of $X$ can be used as a coordinate on
${\cal U}$, at least for $\epsilon$ small enough.
In the coordinate system $(s,y^i)$ the metric takes the form
$$
g_{\mu\nu}dx^\mu dx^\nu= g_{ss}ds^2 + 2g_{is}dy^i ds + g_{ij} dy^i dy^j\ ,
$$
and the set ${\cal O}$ is given by the equation $\{s=0\}$. In
particular $g_{ij}dy^i dy^j$ is the metric induced by
$g_{\mu\nu}dx^\mu dx^\nu$ on ${\cal O}$; by 
construction ${\cal O}$ is spacelike up--to--boundary so that
$g_{ij}dy^i dy^j$ is uniformly non--degenerate and $C^2$ up to the
boundary $\partial \Sigma\cap {\cal U}=\{s=x=0\}$. We also have
$X=X^\mu \partial _\mu = \partial/\partial s$.  $X$ is normal to $\cN$
by hypothesis, which implies that for all vectors $Z^\nu$ tangent to
$\cN$ we have
\begin{equation}
  \label{bc1}
  g_{\mu\nu}X^\mu Z^\nu=0 \quad \Longrightarrow \quad
g_{ss}\Big|_{x=0}=g_{sA}\Big|_{x=0}= 0\ .
\end{equation}
Since the space--time metric is non--degenerate up--to--boundary and
since the closure of the set $\{s=x=0\}$ is compact, there exists a
constant $C$ such that
 \begin{equation}
   \label{bc2}
   C^{-1} \le |g_{sx}|\Big|_{x=0} \le C\ .
 \end{equation}
 Let $\bcU$ denote the set of points in ${\cal U}$ with $x>0$. In
 local coordinates $y^i$ the definition \eq{eq:hdefnew} gives
 \begin{equation}
   h = h_{ij}dy^i dy^j =\Big(g_{ij}+
 \mbox{ $\displaystyle\frac{g_{is}g_{js}}{
     {|g_{ss}|}}$} \Big)dy^i
 dy^j
\label{hdef}\ .
 \end{equation}
We set
\begin{equation}
  \label{eq:vdef}
   V^2\Big|_{\breve{\cal U}} = -g_{ss} > 0\ .
\end{equation}
Clearly $V^2=-g_{\mu\nu}X^\mu X^\nu$, so that $V$ can be extended by
continuity to a function defined on ${\cal U}$, and hence on
$\overline\Sigma$, still denoted by $V$, by setting $V\Big|_{x=0}=0$.
In the $(x,s,v^A)$ coordinate system it holds that
\begin{eqnarray}
 & g_{ss,x}\Big|_{x=0}  = -2\kappa g_{sx} \ . &\label{kdef1}
\end{eqnarray}
In the degenerate case, since the metric is $C^2$ we then have
\begin{equation}
  \label{b1}
  |g_{ss}|\le \hat C x^2\ ,
\end{equation}
for some constant $\hat C$. We wish to show that all curves
$\gamma\subset \Sigma$ that approach the boundary $\{x=0\}$ have
infinite length in the metric $h$. Let us note that $h$ can be written
in the form
\begin{equation}
  \label{he}
  h_{ij}dy^i dy^j = \chi dx^2 +h_{AB}(dv^A+f^A dx)(dv^B+f^B dx)\ ,
\end{equation}
where $f^A$ is a solution of the equation
\begin{equation}
  \label{fe}
  g_{AB}f^B = g_{Ax} + \frac{g_{As}}{|g_{ss}|}(g_{xs}-g_{Bs}f^B)\ ,
\end{equation}
and $\chi$ is given by
\begin{eqnarray}
\nonumber  \chi & = & g_{xx} + \frac{g_{xs}^2}{|g_{ss}|}-h_{AB}f^Af^B
\\ & = & \frac{g_{xs}}{|g_{ss}|}( g_{xs}- g_{Cs}f^C) - g_{Cx}f^C + g_{xx}
\ .
  \label{chi}
\end{eqnarray}
We wish to extract the most singular part of $\chi$ from the equations
above. In order to do that, first note that \eq{bc1} implies that we
have
$$
|g_{As} |\le C x
$$
for some constant $C$, so that one expects the terms
$g_{sx}g_{sx}/|g_{ss}|$ to dominate in \eq{hdef}. This is, roughly
speaking, the case, as can be seen as follows: Equation \eq{fe} leads
to
$$
\Big(|g_{ss}|+ \hat g^{AB}g_{As}g_{Bs}\Big)g_{Cs}f^C= \hat g^{AB}g_{Bs}\Big(
g_{Ax}|g_{ss}|+ g_{xs}g_{As} 
\Big)\ ,
$$
where $\hat g^{AB}$ is the matrix inverse to $ g_{AB}$. It follows
that
\begin{eqnarray*}
g_{AB}f^B  & = &\mbox{ $\displaystyle \frac{g_{As}(g_{sx}-\hat
    g^{CD}g_{Cx}g_{Ds}) }{|g_{ss}|+ \hat
    g^{CD}g_{Cs}g_{Ds}}$}+ g_{Ax}
\\
& = &\mbox{ $\displaystyle \frac{g_{As}g_{sx}}{|g_{ss}|+ \hat
    g^{CD}g_{Cs}g_{Ds}}$}+ \mbox{ $\displaystyle
  O\Big(\frac{x^2}{|g_{ss}|+ \hat
    g^{CD}g_{Cs}g_{Ds}}\Big)$}  + O(1)
\ ,
 \\
\chi &  = & \mbox{ $\displaystyle \frac{(g_{sx}-\hat
    g^{AB}g_{Ax}g_{Bs})^2}{|g_{ss}|+ \hat
    g^{AB}g_{As}g_{Bs}}$} + g_{xx} - \hat
    g^{AB}g_{Ax}g_{Bx}
 \\ & = & \mbox{ $\displaystyle \frac{g_{sx}^2}{|g_{ss}|+ \hat
    g^{AB}g_{As}g_{Bs}}$} +\mbox{ $\displaystyle
   O\Big(\frac{x}{|g_{ss}|+ \hat
    g^{AB}g_{As}g_{Bs}}\Big)$} + O(1)\
\ ,
\end{eqnarray*}
so that there exists a constant $\epsilon > 0$ such that for $x$ small
enough we have
$$
\chi \ge \frac{\epsilon}{x^2}\ .
$$
Consider a curve $[0,1)\ni s \to \gamma(s)\subset \{t=0\} $ such that
$\lim_{s_i\to 1} x(\gamma(s_i))=0$ for some sequence $s_i$. Let $\ell(\sigma)$ denote the
$h$--length of $\gamma([0,\sigma])$; by \eq{he} we have
\begin{eqnarray*}
   \ell(\sigma_i)=\int_0^{\sigma_i} \sqrt{h_{ij}\dot \gamma^i \dot
  \gamma^j } \, ds & \ge & \int_0^{\sigma_i} \sqrt\chi \Big|
  \frac{dx(\gamma(s))}{ds}\Big|  \, ds\\
&  \ge &  \int_0^{\sigma_i} \frac{\sqrt\epsilon}{x(s)} \Big|
  \frac{dx(\gamma(s))}{ds}\Big| \, ds \\
&  \ge & \sqrt\epsilon\Big(|\ln (x(\gamma(\sigma_i))|- |\ln
(x(\gamma(0))|\Big) \ \longrightarrow_{\sigma_i\to
  1} \ \infty
\end{eqnarray*}
(this last inequality requires a not very difficult justification),
which is what had to be established. \qed

An example of the behavior described above is given by the extreme
Reissner--Nordstr\"om space--time. In this case in appropriate
coordinates the metric approaches the standard product metric on the
cylinder $\R \times S^2$ in the asymptotic end constructed as above.

Let us turn our attention to the  case where $\kappa$ has no zeros:

\begin{Proposition}
  \label{pbc2} Let $\Sigma$ be a smooth space--like hypersurface in a
  space--time $(M,g)$ with Killing vector $X$, suppose that $\Sigma$
  is smooth and space--like up to$^{\ref{upto}}$ its topological
  boundary $\pS\equiv\bar\Sigma\setminus \Sigma$, except perhaps at
  those points of $\ps$ at which $X$ vanishes,
  with
  $$g_{\mu\nu}X^\mu X^\nu < 0 \ \mathrm{on}\ \Sigma, \quad
  g_{\mu\nu}X^\mu X^\nu = 0 \ \mathrm{on}\ \partial \Sigma\ .$$
  Then
  every connected component $S$ of $\pS$ which intersects a smooth
  Killing horizon $\cNX$ on which $$\kappa> 0$$
  corresponds to a
  totally geodesic boundary of $(\Sigma,h)$, with $h$ being smooth
  up--to--boundary. Moreover 
  \begin{enumerate}
  \item a doubling\footnote{\label{doubling}See the proof of Theorem \ref{T1}, Section
      \ref{ptt1} below, for an explicit construction of the doubling.}
    of $(\Sigma,h)$ across $S$ leads to a smooth metric on the doubled
    manifold,
  \item with $\sqrt{-g_{\mu\nu}X^\mu X^\nu}$ extending smoothly to
    $-\sqrt{-g_{\mu\nu}X^\mu X^\nu}$ across $S$.
  \end{enumerate}
\end{Proposition}

\noindent\textbf{Remarks:} 1. We note that our proof does \emph{not}
require $\kappa$ to be constant on $\pS$, as long as it has no zeros.
It is nevertheless worth mentioning that in \emph{static} space--times
$\kappa$ is always constant on connected components of $\cN$,
independently of any field equations \cite[Corollary 2.2]{RaczWald2}
(\emph{cf.\/} also \cite[Theorem 8]{CarterlesHouches} in the vacuum
case). Further, independently of staticity, $\kappa$ is constant by
Theorem \ref{tboyer} and by \cite[p.~59]{KayWald} on any connected
component  $\cN_{X}^0$ of $\cN_X$ such that $X$ has zeros on the closure of
$\cN_{X}^0$.

2. When $\Sigma$ is orthogonal to the Killing vector and the
space--time is vacuum or electro--vacuum the result is well known
\cite{Israel:vacuum,Israel:elvac}; in this case our approach seems to
be simpler than elsewhere.  The general result presented here seems to
be new.

3. We emphasize that while the topological boundary $\partial \Sigma$
of $\Sigma$ will typically have corners at those points at which $X$
vanishes, Proposition \ref{pbc2} shows that one can introduce a
differentiable structure on $\bar\Sigma$ such that the $h$--metric
boundary of $\Sigma$ will be a smooth submanifold of $\bar\Sigma$.  .

\medskip

\Proof We shall treat the case in which $X$ has no zeros on $\partial
\Sigma$ separately:

\begin{Lemma}
   \label{lnoz} Under the hypotheses of Proposition \ref{pbc2},
   assume further that $X$ has no zeros on some open connected set
   ${\cal O}\subset \partial \Sigma $. Then the conclusion of
   Proposition \ref{pbc2} holds near ${\cal O}$.
\end{Lemma}

\proof As $X$ has no zeros on ${\cal O}$ we can construct a coordinate
system $(s,x,v^A)$ in a neighborhood of ${\cal O}$ as in the proof of
Proposition \ref{pbc1}. From Eq.\ \eq{kdef1} we have
$$
V^2 = 2\tk x + O(x^2)\ , \qquad \tk = \kappa g_{sx}\Big|_{x=0} \ .
$$
Let $x=w^2$, it holds that
\begin{eqnarray}
\nonumber
  h_{ij}dy^i dy^j & = & 4\Big( g_{sx}^2 \Big(\frac{w}{V}\Big)^2 +
  w^2g_{xx}\Big)\,dw^2 + 4w \Big(  g_{xA}  +g_{sx} \frac{g_{sA}}{w^2}
  \Big(\frac{w}{V}\Big)^2  \Big)\, dv^Adw
\\ & &  + \Big(  g_{AB}  +g_{sA} \frac{g_{sB}}{w^2}
  \Big(\frac{w}{V}\Big)^2  \Big)\, dv^Adv^B
\ .  \label{bc10}
\end{eqnarray}
Note that all the functions appearing in \eq{bc10} are bounded, and
that $g_{AB}$, $g_{xx}$ and $g_{sx}$ are smooth functions of
$(w^2,v^A)$ up to  the boundary $\{w=0\}$. We also have
$g_{sA}=x g_A$ for some functions $g_A$ which are smooth in $(x,v^A)$,
so that $\frac{g_{sA}}{w^2}$ is again a smooth function of $(w^2,v^A)$
up to the boundary $\{w=0\}$. Similarly $g_{ss}=-x g$
for some function $g$ which is smooth up--to--boundary in $(x,v^A)$,
with $g|_{x=0} = 2\tk \ne 0$. It follows that ${w}/{V}=
{1}/{\sqrt{g}}$ is a a smooth function of $(w^2,v^A)$, up to
 $\{w=0\}$. Equations \eq{bc1}--\eq{bc2} further show that the
metric is uniformly non--degenerate up--to--boundary. The fact that
the boundary is totally geodesic follows immediately from the
definition of the extrinsic curvature and from the equality
$$
\frac{\partial \phi (w^2,v^A)}{\partial w}\Big|_{w=0} =
2w\frac{\partial \phi (x,v^A)}{\partial x}\Big|_{x=0} = 0\ ,
$$
for any differentiable function $\phi (x,v^A)$.

To justify the claims about the doubled
manifold$^{\mbox{\scriptsize\ref{doubling}}}$, recall that the double
is constructed by allowing $w$ to take negative values and by
extending the relevant functions via $f(w)=f(-w)$. Thus $w^2$ is a
smooth function on the doubled manifold in a neighborhood of $S$, and
we have $$4w \Big( g_{xA} +g_{sx} \frac{g_{sA}}{w^2}
\Big(\frac{w}{V}\Big)^2 \Big)\, dv^Adw = 2 \Big( g_{xA} +g_{sx}
\frac{g_{sA}}{w^2} \Big(\frac{w}{V}\Big)^2 \Big)\, dv^Ad(w^2)\ ,$$
which is a smooth tensor field on the doubled manifold from what has
been said above. The remaining coordinate components of $h$ are
obviously smoothly extendible to the double. Similarly $V/w$ is a
smooth scalar field when extended by an even function of $w$ on the
double, and our claims follow.  \qed

Proposition \ref{pbc2} follows immediately from Lemma \ref{lnoz}
and from the following:

\begin{Lemma}
   \label{pbc3} Under the hypotheses of Proposition \ref{pbc2},
   assume instead that $X$ vanishes at a point $p\in\partial \Sigma $.
   Then there exists a neighborhood ${\cal O}\subset \partial \Sigma $
   of $p$ such that the conclusion of Proposition \ref{pbc2} holds
   near ${\cal O}$.
\end{Lemma}

{\noindent{\bf Remark:}\ } We emphasize that we do \emph{do
not} assume that $X$ vanishes throughout $\partial \Sigma$.

\medskip

\proof By Theorem \ref{tboyer} there exists a neighborhood $\cV$ of
$p$ such that the set $\{p\, |\, X(p)=0 \}\cap \cV$ is a smooth,
embedded, space--like, two--dimensional submanifold of $M$. Passing to
a subset of $\cV$ if necessary, $\cV$ can be covered by a
R{\'a}cz--Wald--Walker coordinate system\footnote{The coordinates
  $(u,v,x^a)$ here correspond to the coordinates $(U,-V,x^\alpha)$ of
  \cite{RaczWald1}; we also use a different normalization of the
  Killing vector $X$.}  \cite{RaczWald1,Walker} $(u,v,x^a)$, with
$x^a$ being coordinates on $\{p\, |\, X(p)=0 \}\cap \cV$, and with $u$
and $v$ covering the set $|uv|<\epsilon$, $|u|<\epsilon$,
$|v|<\epsilon$, for some $\epsilon>0$. In this coordinate system the
space--time metric $g$ takes the form
\begin{equation}
  \label{eq:rw}
  g=2Gdu\, dv + 2 v H_adx^adu + g_{ab}dx^adx^b \ ,
\end{equation}
with $G$, $h_a$, $g_{ab}$ being smooth functions of
$(uv,x^a)$. Here $g_{ab}$ is a two by two strictly positive matrix,
and $G$ satisfies
$$
C^{-1}\le G \le C\ ,$$
for some constant $C$. Orientations have
been chosen so that $\Sigma\cap\cV\subset \{u>0, v>0\}$, which implies
$\partial\Sigma\cap\cV\subset \{u=0\}\cup\{ v>0\}$. In this coordinate
system the Killing vector field $X$ takes the following simple form:
\begin{equation}
  \label{eq:rwx}
  X=u\frac{\partial}{\partial u} - v\frac{\partial}{\partial v}\ .
\end{equation}
Since the vector field $\partial/\partial v$ is light-like throughout
$\cV$ it is transverse to $\Sigma$, hence in a neighborhood of any of
its points $\Sigma\cap \cV$ can be described as
$$  \Sigma\cap \cV=\{v=\phi(u,x^a)\}\ ,$$
for some smooth function $\phi$. In this coordinate system the metric
$\gamma$ induced by $g$ on $\Sigma$ takes the form
\begin{equation}
  \label{eq:gamma}
  \gamma=2Gdu\, d\phi + 2 \phi H_a dx^a du + g_{ab}dx^adx^b\ .
\end{equation}
The hypothesis that $\Sigma$ is spacelike up to boundary implies that
\begin{equation}
  \label{eq:uco}
 0 < \frac{\partial \phi}{\partial u}\ .
\end{equation}
We further have
\begin{equation}\label{xf}
  g(X,\cdot)= G(u\, dv - v\, du) + uv H_a dx^a\ ,
\end{equation}
so that the orbit space metric $h$, as defined by Equation
\eq{eq:hdefnew}, reads
\begin{equation}
  \label{eq:hag}
  h = \frac{G}{2uv}(d(u\phi))^2 + 2 H_a dx^ad(u\phi) +  \Big( g_{ab}+
  \frac{u\phi}{2G } H_a H_b \Big)dx^adx^b\ .
\end{equation}
Define a function $w$ on $\Sigma\cap \cV$ by the formula
$$
w^2 = \phi u\ .$$
We have $\phi u=(v u )|_\Sigma>0$ so that $w$ is
well defined and takes real values. Moreover the derivative
$$ \frac{\partial (w^2)}{\partial u} = u \frac{\partial \phi}{\partial
  u}+  \phi
$$
is strictly positive in virtue of Equation \eq{eq:uco} and of the
inequalities $\phi>0$, $u|_\Sigma >0$. This shows that $(w,x^a)$ can
be used as coordinates on $\Sigma\cap\cV$. In this coordinate system
we have
\begin{equation}
  \label{eq:youpi}
  h = 2{G}(dw)^2 + 4 wH_a dx^adw +   \Big( g_{ab}+
  \frac{w^2}{2G } H_a H_b \Big)dx^adx^b\ .
\end{equation}
We note that the functions $G$, $H_a$ and $g_{ab}$ appearing in
this last equation are smooth functions of $(w^2,x^a)$. The set $\cO$
is defined now as $\partial\Sigma\cap\cV$. The remaining claims follow
now as in the proof of Lemma \ref{lnoz}. \qed

\section{A proof based on the Komar identity}
\label{Komar}
In the analysis below we shall
need the Vishveshwara--Carter Lemma:

\begin{Lemma}[Vishveshwara--Carter Lemma \cite{Vishveshwara,CarterJMP}]
  \label{lcvnew}
  Let $(M,g)$ be a smooth space--time with Killing vector $X$
  satisfying the staticity condition \eq{eq:st}, define $\hat {\cal
    N}$ to be the boundary of the set $ \{g_{\alpha\beta}X^\alpha
  X^\beta <0\}$, set
 $${\cal N}\equiv \hat {\cal N}\cap \{X^\alpha\ne 0\}.$$ Then
${\cal N}$ is a smooth null hypersurface, with
the Killing vector field $X$ normal to  ${\cal N}$. \ptc{there
is a problem here, $\cal N$ could fail to be embedded; a
corrected version of the Lemma can be found in the Addendum.
The corrected, weaker version is not sufficient for the problem
at hand, but the issue is solved in~\cite{ChGstaticity} under
supplementary global hypotheses.}
\end{Lemma}

{\noindent{\bf Remark:}\ } 
We emphasize that it is not assumed here that  ${\cal N}$ satisfies
some non--degeneracy conditions.

\medskip

\proof Let $p\in \cN$; from the staticity condition \eq{eq:st} and
from the Frobenius theorem \cite[Section 9.1]{Hicks} there exists a
neighborhood $\cal O$ of $p$, with $X$ nowhere vanishing on ${\cal
  O}$, foliated by a family of smooth hypersurfaces $\Sigma_\tau$
which are normal to $X$. By passing to a subset of  $\cal O$ if
necessary we may without loss of generality suppose that $\cal O$ has
compact closure. The staticity condition \eq{eq:st} can be
rewritten as
$$
2 \nabla_{[\mu} X^\alpha X_{\nu]} =  X^\alpha \nabla_{[\mu} X_{\nu]}\
,
$$
which after a contraction with $X_\alpha$ gives
\begin{equation}
  \label{eq:prop}
   \nabla_{[\mu}W X_{\nu]} = W \nabla_{[\mu} X_{\nu]}\
, \qquad W\equiv X_\alpha X^\alpha\ .
\end{equation}
On ${\cal O}\cap \{W<0\}$ eq.\ \eq{eq:prop} takes the form
\begin{equation}
  \label{eq:prop1}
   \nabla_{[\mu}\Big(\ln(-W)\Big) X_{\nu]} =  \nabla_{[\mu} X_{\nu]}
\ .
\end{equation}
Let $\ell_\mu$ be any smooth covector field on ${\cal O}$ such that
$\ell_\mu X^\mu=1$ and let $Z^\mu$ be any smooth vector field tangent
to a leaf $\Sigma_\tau$ such that $\Sigma_\tau\cap\{W<0\}\ne \emptyset
$; at points at which \eq{eq:prop1} holds contraction of this equation
with $Z^\mu \ell ^\nu$ gives
\begin{equation}
  \label{eq:prop2}
  Z^\mu \nabla_{\mu}\ln(-W)  = 2 Z^\mu \ell ^\nu \nabla_{[\mu} X_{\nu]}
\ .
\end{equation}
The right--hand--side of the above equation is uniformly bounded on
$\Sigma_\tau$, so that $\ln(-W)$ has uniformly bounded
gradient on   $\Sigma_\tau\cap
\{W<0\}$. In particular  $\ln(-W)$ is uniformly bounded
 on   $\Sigma_\tau\cap
\{W<0\}$, which is only possible if $\Sigma_\tau\cap
\{W=0\}=\emptyset$. This shows that if $\Sigma_{\tau_0}\cap
\{W=0\}\ne\emptyset$ then $W\equiv 0$ on $\Sigma_{\tau_0}$. Hence
$\{W=0\}\cap {\cal O}$ is a union of leaves of the  $\Sigma_\tau$
foliation. In particular each connected component of ${\cal N}\cap
{\cal O}$ coincides  with a leaf of the  $\Sigma_\tau$
foliation, hence is a smooth hypersurface normal to $X$. \qed

Non--existence of black holes with \emph{all} connected components of
the event horizon degenerate, or with empty $\partial \Sigma$, can be
established as follows: Suppose, as in Theorem \ref{T1}, that the
space--time is static, and that $\bar \Sigma$ is the union of a finite
number of asymptotically flat ends and of a compact interior. We have
the Komar identity
 \begin{eqnarray}
   \nonumber
   m_K & = & \frac{1}{8\pi} \int _{S_\infty} \nabla^\mu X^\nu
   dS_{\mu\nu}
 \\ & = &
   \frac{1}{8\pi} \Big\{\frac{1}{2}\int_\Sigma   \nabla_\mu\nabla^\mu X^\nu
   dS_{\nu} +\int _{\partial\Sigma} \nabla^\mu X^\nu dS_{\mu\nu}\Big\}
\nonumber \\    & = &
   \frac{1}{8\pi} \int _{\partial\Sigma} \nabla^\mu X^\nu dS_{\mu\nu}\
   .
   \label{I1new}
 \end{eqnarray} Here  $m_K$ is the sum of the
 Komar masses of the asymptotically flat ends, as defined by the first
 line of Equation \eq{I1new}, $X^\mu$ is the Killing vector field
 which asymptotes to $\partial/\partial t$ in the asymptotically flat
 ends, ${S_\infty}$ is the ``union of spheres at infinity'' in the
 asymptotically flat ends, and we have used the equation $
 \nabla_\mu\nabla^\mu X^\nu =0 $ that holds for a Killing vector field
 in vacuum.  A theorem of Beig \cite{BeigKomar} (\emph{cf.\/} also
 \cite{Chremark,ashtekar:magnon:conserved}) shows that the Komar mass
 of each static asymptotically vacuum end $(\Sext, g|_{\Sext})$
 coincides with its ADM mass, so that $m_K$ coincides with the sum of
 the ADM masses of the asymptotically flat ends.  Boyer's Theorem
 \ref{tboyer} shows that $X$ is nowhere vanishing on $\partial
 \Sigma$. Further, the Vishveshwara--Carter Lemma \ref{lcvnew}
 \ptc{here $\partial \Sigma$ is assumed to be a compact
 embedded topological manifold, so the embededness problem
 in Lemma~\ref{lcvnew} does not arise by hypothesis}
 and staticity imply that $\partial \Sigma$ is a subset of the
 Killing horizon, in particular (slightly deforming $\Sigma$ if
 necessary) $\partial \Sigma$ is a smooth submanifold of
 $\overline\Sigma$. Those facts and an easy calculation show
 (\emph{cf., e.g.}, \cite{BCH}) that
\begin{equation}
  \frac{1}{8\pi}\int _{\partial\Sigma} \nabla^\mu X^\nu dS_{\mu\nu}
   = \frac{1}{4\pi}\sum
   \kappa_i    A_i = 0\ . \label{I2new}
\end{equation}
Here the $\kappa_i$'s are the surface gravities of the connected
components of the event horizon (which vanish by hypothesis), and the
$A_i$'s are the areas of the corresponding components of $\partial
\Sigma$; in the last line of \eq{I2new} the sum is over an empty index
set if $\partial \Sigma$ is empty. If $\partial \Sigma=\emptyset$ it
follows from the rigid positive energy theorem proved in
\cite{ChBeig1} that $\Sigma$ can be embedded into Minkowski
space--time so that the metric on $\Sigma$ is the pull--back of the
Minkowski space--time metric, with the extrinsic curvature tensor of
$\Sigma$ taking the appropriate values corresponding to the embedding.
(Further the image of $\Sigma$ has to be a Cauchy surface in Minkowski
space--time \cite{ChBeig1}; in particular the maximal globally
hyperbolic vacuum development of the initial data induced by $g$ on
$\Sigma$ is the Minkowski space--time.)  If $\partial
\Sigma\ne\emptyset$ the positive mass theorem with marginally trapped
boundary \cite{Herzlich:mass} gives a contradiction. This proves that
the case in which all components of the boundary of $\Sigma$ meet
degenerate Killing horizons cannot occur under the hypotheses of
Theorem \ref{T1}.

It is of interest to enquire how to modify this argument if stationary
asymptotically flat space--times are considered which do not
necessarily satisfy the staticity condition. Clearly the Komar
identity (\ref{I1new}) still holds. If $\partial \Sigma=\emptyset$,
the fact that $(\Sigma,g,K)$ can be appropriately embedded in
Minkowski space--time follows as above, whenever $\Sigma$ is the union
of a finite number of asymptotically flat regions and of a compact
set. This is the classical Lichnerowicz
theorem$^{\mbox{\scriptsize\ref{lichne}}}$. Now, if $\partial
\Sigma\ne\emptyset$, the Vishveshwara--Carter lemma does not apply any
more, so the Killing vector does not need to be tangent to the
generators of $\cN$.  Further a degenerate non--static horizon could
be non--differentiable.  We note that in this case one can proceed as
follows: Let $(M,g)$ be an asymptotically flat solution of the vacuum
Einstein equations with Killing vector field $X$.  Suppose, as in
Theorem \ref{T1}, that $M$ contains a connected space-like
hypersurface $\Sigma$ the closure $\bar \Sigma$ of which is the union
of a finite number of asymptotically flat ends and of a compact
interior, such that the topological boundary $\partial \Sigma\equiv
\overline{\Sigma}\setminus \Sigma$ of $\Sigma$ is a nonempty
topological manifold.  Assume that the Killing vector $X$ is timelike
future directed in all the asymptotic regions. Suppose further that
$\partial \Sigma$ is a subset of a degenerate Killing horizon $\cal
N$. This is certainly a supplementary hypothesis, as compared to the
static case, which can be interpreted as the hypothesis that \emph{all
  Killing horizons are non--rotating}. If one knew that the horizon is
differentiable one could obtain the equality (\ref{I2new}), and
conclude as before that no such space--times exist. Let us show that
the required differentiability must hold:
\begin{Proposition}\label{Pkilling}
  A Killing horizon is a \emph{locally achronal} hypersurface of at
  least $C^1$ differentiability class. More precisely, the
  intersection of a connected component of a Killing horizon with any
  connected, globally hyperbolic open set $\cO$ is a $C^1$ achronal
  hypersurface in $\cO$.
\end{Proposition}

{\noindent{\bf Remark:}\ } We note that a Killing horizon must
be $C^\infty$ (or as differentiable as the metric allows) in a
neighborhood of any point thereof for which $\kappa\ne 0$ --
this follows immediately from the fact that at such points
$g_{\mu\nu}X^\mu X^\nu$ has non--zero gradient. Thus the only
problematic points as far as differentiability is concerned are
those at which $\kappa$ vanishes.

\medskip

\proof Let us start by showing that a Killing horizon $\cN$ is necessarily
\emph{locally} achronal, that is, for every $p\in \cN$ there exists a
neighborhood $\cO$ such that
$$
\forall \ q\in \cN\cap\cO \qquad I(p;\cO)\cap I(q;\cO) = \emptyset\ .
$$
Indeed, let $\cO$ be any globally hyperbolic neighborhood $\cO$ of
$p$ such that $\cN\cap\cO$ is connected; changing time orientation if
necessary connectedness of $\cN\cap\cO$ shows that for all $q\in
\cN\cap\cO$ and for all future directed timelike vectors $T$ in $T_qM$
we have
\begin{equation}
  \label{eq:product}
  g(T,X)< 0\ .
\end{equation}
Let $q\in I(p;\cO)$, by global hyperbolicity there exists a timelike
geodesic segment $\gamma\subset\cO$ with $q$ and $p$ as endpoints. We have
$$\frac{d g(\dot \gamma,X)}{ds}= 0 \ ,
$$
which together with \eq{eq:product} shows that $\gamma$ can intersect
$\cN$ only at p.

We have thus shown that $\cN$ is necessarily a \emph{locally} achronal
null hypersurface generated by the null geodesics tangent to $X$, such
that every point on $\cal N$ is an interior point of a generator.
Hence, by well known properties of such hypersurfaces, $\cal N$ has to
be of $C^1$ differentiability class (\emph{cf., e.g.\/}
\cite{Chendpoints} for a simple proof). \qed

\section{Considerations global in space}
\label{ptt1}
We shall need the following, essentially obvious, result. For future
reference, we state it in a context more general than the vacuum
Einstein equations:

\begin{Lemma}
  \label{lfeq} Suppose that $(M,g)$ is static, and suppose that the
  couple $(\hat h, \hat V)$, where $\hat h$ is the metric induced on
  the hypersurfaces orthogonal to $X$ and $-\hat V^2$ is the square of
  the Lorentzian norm of $X$ on those hypersurfaces, satisfies some
  coordinate--independent system of equations. Then the orbit
  space--metric $h$ together with  the function $V$ (such that $-
  V^2$ is the square of the Lorentzian norm of $X$ on $\Sigma$)
  satisfies the same system of equations.
\end{Lemma}

\proof For any point $p\in\Sigma$ there exists a neighborhood $\bcU $
on which we can construct coordinates as in the beginning of the proof
of Proposition \ref{pbc1}, with the difference that we are not
assuming that we are near the boundary, so that the set ${\cal
  O}\subset \Sigma$ considered there is now some neighborhood of $p$ in
$\Sigma$. Passing to an appropriate smaller subset of ${\cal O}$ and
then decreasing $\epsilon$ if necessary, where $\epsilon$ is as in the
proof of Proposition \ref{pbc1}, staticity and the Frobenius theorem
imply that there exists a function $t\in C^{\infty}(\bcU)$ ($\bcU$
again as in the proof of Proposition \ref{pbc1}) such that the metric
takes the form
\begin{equation}
  \label{bc3}
  ds^2 = -\hat V^2dt^2 + \hat h \ , 
\end{equation}
with $X=\partial/\partial t$. The level sets of $t$ are by definition
normal to the Killing vector field which is time-like on $\bcU$, so
that $\hat h$ is Riemannian. This time function $t$ is defined uniquely up
to a constant. On $\bcU$ we have $X=\partial/\partial s =
\partial/\partial t$, which implies
\begin{eqnarray}
  & t=s+f(y^i)\ , \quad  dt = ds+  \mbox{
    $\displaystyle\frac{g_{si}}{
      {g_{ss}}}$}dy^i\ ,
\label{tdefagain}
  & \\
& \hat V^2\Big|_{\breve{\cal U}} = -g_{ss} = V^2> 0\ ,
\nonumber
&\\
& \hat h = \hat h_{ij}dy^i dy^j =\Big(g_{ij}+
 \mbox{ $\displaystyle\frac{g_{is}g_{js}}{
     {|g_{ss}|}}$} \Big)dy^i
 dy^j = h_{ij}dy^i dy^j
\label{hdefagain}\ , &
\end{eqnarray}
and the result follows. \qed

In the proof of Theorem \ref{T1} we shall need the following version of the
positive energy theorem, proved$^{\mbox{\scriptsize \ref{geroch}}}$ in
\cite{BartnikChrusciel1}:

\begin{Theorem}
  \label{tpet}
  Let $(\Sigma,h)$ be a smooth complete Riemannian manifold with an
  asymptotically flat end $\Sigma_\ext$ and with positive scalar
  curvature. If the Ricci scalar is integrable\footnote{If the Ricci
    scalar is not integrable the conclusion still holds with
    $m=\infty$.} on $\Sigma_\ext$ then the ADM mass of $\Sigma_\ext$
  satisfies
$$m\ge 0\ .$$
Moreover, if the equality is attained, then there exists a
diffeomorphism $\psi:\Sigma\to\R^3$  such that $h$ is the pull--back
by $\psi$ of the standard Euclidean metric $\delta$ on $\R^3$.
\end{Theorem}

We emphasize that in the result above $\Sigma$ can have an arbitrary
number (perhaps infinite) of asymptotic ends, and that \emph{no
  hypotheses} are made on the asymptotic behavior of the metric in
those ends except that the metric $h$ is complete (and except, of course,
that at least one of the ends is asymptotically flat so that its ADM mass is
well defined). More general results, allowing for non--vanishing extrinsic
curvature of the initial data hypersurface, poor differentiability of the metric,
and boundaries, can be found in \cite{BartnikChrusciel1}.

\medskip

We are ready now to pass to the

\medskip

\noindent\textsc{Proof of Theorem \ref{T1}:} Consider the
manifold $\Sigma$ equipped with the metric $h$ given in local
coordinates by Equation \eq{hdef}.  By Lemma \ref{lcvnew}
 \ptc{here $\partial \Sigma$ is assumed to be a compact embedded topological manifold,
 so the embededness problem arising in Lemma~\ref{lcvnew} does not arise by hypothesis}
the boundary $\partial \Sigma$ is a subset of the closure of
the Killing horizon $\cN$, which is a smooth submanifold of $M$
by that same lemma except at points at which $X$ vanishes.
Slightly deforming $\Sigma$ in space--time if necessary we can
without loss of generality assume that $\partial\Sigma$ is a
smooth sub-manifold both of the space--time and of $\overline
\Sigma$, except perhaps at those points at which $X$ vanishes.
Further deforming $\Sigma$ if necessary we may assume that the
metric on $\overline \Sigma$ is \emph{spacelike up to
boundary}. We can thus use Propositions \ref{pbc1} and
\ref{pbc2} to conclude that the pair $(\Sigma,h)$ is a complete
Riemannian manifold with compact boundary and with at least one
asymptotically flat end $\Sext$.  Let us denote by $\Snd$ the
collection of all those components of the boundary of $\Sigma$
which correspond to non--degenerate components of the event
horizon of the black hole; by Propositions \ref{pbc1} and
\ref{pbc2} the metric boundary of $(\Sigma,h)$ is $\Snd$; it is
compact and totally geodesic.  (We note that the case where
there are only degenerate horizons has already been excluded in
Section \ref{Komar}.) By Lemma \ref{lfeq} the metric $h$ and
the function $V$ satisfy the equations (\emph{cf., e.g.},
\cite{bunting:masood})
\begin{eqnarray}
  \label{bc11}
&   \Delta_h V = 0 \ , &
\\ &   \label{bc12}
R_{ij}= V^{-1} D_i D_j V
\ , &
\end{eqnarray}
where $\Delta_h$ is the Laplace--Beltrami operator of the metric $h$
and $R_{ij}$ is its Ricci tensor. Following \cite{bunting:masood}, set
 \begin{eqnarray}
   &\nonumber
\Sigma_+ = \Sigma, \qquad h_+ = \Big(\frac{1+V}{2}\Big)^4 h\ ,
& \\ &\nonumber
\Sigma_- = \Sigma\cup \{\Lambda_i\}, \qquad h_- =
\Big(\frac{1-V}{2}\Big)^4 h\ ,
& \\ &\label{hh}
\hS = 
\Sigma_+ \cup \Sigma_-\cup \Snd
\ , \qquad \hat
h\Big|_{\Sigma_+}=h_+\ , \quad \qquad \hat
h\Big|_{\Sigma_-}=h_-
\ .&
 \end{eqnarray}
 Here $\Sigma\cup \{\Lambda_i\}$ denotes a one point compactification of
 all the asymptotically flat regions 
 of $\Sigma$ (with a point $\Lambda_i$ for each asymptotically flat
 region). By the results of \cite{BeigSimon2} the metric $h_-$ can be
 extended across the ``points at infinity'' $\Lambda_i$ in a smooth
 (even analytic) way to a Riemannian metric on $\Sigma_-$ still
 denoted by $h_-$.
The topological and differentiable structure of $\hS$ are defined
 through the 
gluing of 
$\overline\Sigma_+\equiv\Sigma_+\cup \Snd $ with
$\overline\Sigma_-\equiv\Sigma_-\cup \Snd $ by identifying $\Snd$,
considered as a subset of $\overline\Sigma_+$, with a second copy of
$\Snd$, considered as a subset of $\overline\Sigma_-$, using the
identity map. Proposition \ref{pbc2} shows that the metric $\hat
h$ defined in \eq{hh} can be extended by continuity to a smooth metric
on $\hS$.  Note that near degenerate components of the event horizon,
if any, we have $\frac{1\pm V}{2}\approx \frac{1}{2}$, which implies
that the conclusion of Proposition \ref{pbc1} still holds for both
$h_+$ and $h_-$, thus the ends of $\hS$ corresponding to degenerate
components of the event horizon are complete both for $h_+$ and $h_-$.
It follows that $(\hS,\hat h)$ is a complete Riemannian manifold
without boundary.  On $\Sext$ we have $\frac{1+ V}{2}\approx 1$, so
that $(\Sext,\hat h|_{\Sext})$ is an asymptotically flat end.  A
theorem of Beig \cite{BeigKomar} (\emph{cf.\/} also
\cite{Chremark,ashtekar:magnon:conserved}) shows that the Komar mass
of a static asymptotically vacuum end $(\Sext, h|_{\Sext})$ coincides
with its ADM mass $m$, which gives
$$
V = 1 -\frac{m}{r}+ O(r^{-2})\ ,
$$ and which implies
that the ADM mass of $(\Sext,\hat h|_{\Sext})$ vanishes. 
Equation \eq{bc11} 
and the behavior of the Ricci scalar under conformal rescalings shows
that the Ricci scalar $\hat R$ of $\hat h$ is non--negative (\emph{cf.,
  e.g.}, \cite[Eq.\ 18]{bunting:masood}). The rigidity part of Theorem
\ref{tpet} shows that there exists a diffeomorphism $\psi:\hS\to\R^3$
such that $\hat h$ is the pull--back by $\psi$ of the standard
Euclidean metric $\delta$ on $\R^3$. This shows that $\hat h$, and
hence also $h$, are conformally flat.

Let $S$ be any connected component of $\Snd$ considered as a subset of
$\hS$, then $\psi(S)$ is an embedded sub-manifold of $\R^3$. $S$ is
totally geodesic with respect to the metric $h$, so that from the
transformation formulae for the extrinsic curvature under conformal
rescalings together with the constancy of the surface gravity on $S$
it follows that $\psi(S)$ has constant mean curvature with respect to
the flat metric on $\R^3$ (\emph{cf., e.g.}, \cite[Lemma
4]{bunting:masood}).  By Alexandrov's theorem \cite[Theorem
2.6]{EellsRatto} the manifold $\psi(S)$ is a coordinate sphere.

Suppose that $\Snd$ had more than one component. Then $ \psi(\Snd)$
would consist of a finite union of disjoint coordinate spheres, and
$\hS\setminus \overline\Sigma_+\approx \R^3\setminus
\overline{\psi(\Sigma_+)}\approx\Sigma_-$ would be a union of disjoint
balls, in particular $\Sigma_-$ would not be connected, which
contradicts connectedness of $\Sigma_-\approx \Sigma$. It follows that
$\Snd$ has precisely one connected component, with $\Sigma$
diffeomorphic to $\R^3\setminus B(0,R)$, so that $(\Sigma,h)$ has only
one asymptotic end and a connected compact boundary lying at finite
$h$--distance.  This establishes non--existence of degenerate event
horizons in static vacuum space--times.

To finish the proof, that $h$ has to be the metric induced on the standard
$t=\textrm{const}$ slices of the Schwarzschild--Kruszkal--Szekeres
space--time, we follow an idea of Herzlich (private communication).
On $\R^3\setminus B(0,R)$ consider the space--Schwarzschild metric of
mass $2R$ (\emph{cf.\/} Equation (\ref{eq:confs})):
$$
\tilde h = \Omega_0^4 \delta , \qquad \Omega_0 \equiv 1+\frac{R}{r}\ .
$$
The sphere $S(0,R)$ is a totally geodesic surface with respect to
$\tilde h$. Let us still denote by $h$ the pull--back by $\psi^{-1}$
of $h$ to $\R^3\setminus B(0,R)$, and use $V$ for $V\circ \psi^{-1}$;
we have
$$
h = \Big(\frac{1+V}{2}\Big)^{-4} \delta = \Big(\frac{1+V}{2}\Big)^{-4}
\Omega_0^{-4} \tilde h = (1+\phi)^4 \tilde h\ ,
$$
with $\phi$ defined by the last equality above, $\phi = 2/((1+V)
\Omega_0) -1$. Both $h$ and $\tilde h$ have vanishing scalar
curvature, so that the transformation law for the Ricci scalar under
conformal transformations shows that $\phi$ is $\tilde h$--harmonic:
$$
\Delta_{\tilde h} \phi = 0 \ .
$$
Here $\Delta_{\tilde h}$ is the Laplace--Beltrami operator of the
metric $\tilde h$. Now $\psi(S)=S(0,R)$ is minimal for both $h$ and $\tilde
h$, and the transformation law for the mean extrinsic curvature shows
that $\phi$ has vanishing Neumann data:
\begin{equation}
  \label{neum}
\frac{\partial \phi}{\partial n}\Big|_{S(0,R)}= 0\ .
\end{equation}
Here it does not matter whether the partial derivative in the normal
direction is taken with respect to the metric $h$ or with respect to
the metric $\tilde h$.  It follows that
\begin{eqnarray*}
  \int_\Sigma |d \phi|_{\tilde h}^2 d^3\mu _{\tilde h} & = &
\int_{S_{\infty}}\phi \frac{\partial \phi}{\partial n} d^2\mu _{\tilde h}
- \int _ S
  \phi \frac{\partial \phi}{\partial n} d^2\mu _{\tilde h} \\
& = & 0 \ .
\end{eqnarray*}
Here the integral over the sphere at infinity $S_{\infty}$ vanishes by
the well known asymptotic behavior of harmonic functions on
asymptotically flat ends,
$$
\phi = O(r^{-1}), \qquad \frac{\partial \phi}{\partial x^i}
 = O(r^{-2})\ ,
 $$
 while the integral over $S$ vanishes because of the vanishing
 Neumann data \eq{neum}. We thus have
$$\phi\equiv 0\ ,
$$
hence $h=\tilde h$, which immediately shows that the metric
$-V^2dt^2+h$ on $\R\times\Sigma$ is the Schwarzschild metric.

Consider any neighborhood $\cU$ of $\Sigma$ diffeomorphic to an open
interval times $\Sigma$; the set $\cU$ is simply connected because
$\Sigma$ has been shown to be simply connected. Let $\alpha$ be the
one--form
$$
\alpha= \frac{\hX_\mu dx^\mu}{\hX_\nu \hX^\nu}\ ;
$$
Equation \eq{eq:prop} shows that $\alpha$ is closed, and
simple--connectedness of $\cU$ implies existence of a function $t\in
C^\infty (\cU)$ such that $\alpha=dt$.  As in the proof of Lemma
\ref{lfeq} (\emph{cf.\/} Eq.\ \eq{tdefagain}) there exists a function
$f:\Sigma\to\R$ such that
\begin{equation}
  \label{eq:embed}
  t=s+f\ ,
\end{equation}
except that now the function $f$ is defined globally on $\Sigma$,
while in Lemma \ref{lfeq} $f$ was only locally defined on
appropriately small open sets.  Here $s$ denotes the coordinate along
the (perhaps only locally defined) orbits of the Killing vector field.
Passing to a subset of $\cU$ if necessary we may assume that every
orbit of $X$ in $\cU$ intersects $\Sigma$ precisely once. We can then
extend $f$ to a function on $\cU$ by requiring that $X(f)=0$. As the
metric $-V^2dt^2+h$ has already been shown to be the Schwarzschild
metric, Equation \ref{eq:embed} provides now the required embedding of
$\cU$ into an open subset of the Schwarzschild space--time. \qed

\section{Considerations global in space--time}
\label{ptt1.1}

While the considerations of the previous section had an essentially
Riemannian character, here we return to an analysis of the geometry of
the space--times under consideration.  Let us start with the

\medskip

{\noindent\textsc{Proof of Corollary \ref{c1}:}}
Suppose, first, that each orbit of the flow $\phi_t$ of $X$ through
$\ks$ intersects $\Sigma$ precisely once. Because $X$ is transverse to
$\Sigma$, it follows that the Killing development $\ks$ of $\Sigma$ is
diffeomorphic to $\R\times \Sigma$ in the standard manner. In this
case the set $\cU$ considered at the end of the proof of Theorem
\ref{T1} can be taken to be equal to $\ks$.  Consider the map
$$\ks\approx\R\times \Sigma\ni
(s,q)\to\Psi(t,q)=(t=s+f(q),q)\in M^{\mathrm{Schw}}\ ,
$$ where $f$ is as at the end of the proof of Theorem \ref{T1}. $\Psi$
is injective because $\Psi(0,q)$ is injective from $\Sigma$ to
$M^{\mathrm{Schw}}$, and because all Killing orbits intersect both
$\Sigma$ and its image precisely once.  $\Psi$ is a local
diffeomorphism by construction, hence an embedding. It follows from
the proof of Theorem \ref{T1} and from the formula above that $\Psi$
is surjective, thus $\Psi$ is a diffeomorphism. This establishes the
second part of the Corollary when all the orbits of $X$ intersect
$\Sigma$ at most once.  In particular our claims follow under the
assumption that $\Sigma$ is achronal in $\doc$.

Suppose, next, that some orbits of $X$ intersect $\Sigma$ more than
once. In that case we can replace $\ks$ by its
Hawking covering $\hat \ks$ as defined in \cite{HHM} in the argument
we just made, equipped with the metric $\pi^*g$, where $\pi:\hat
\ks\to\ks$ is the covering map. In this covering one of the connected
components of the pre-image of $\Sigma$ under $\pi$ will be
homeomorphic to $\Sigma$ \cite{HHM}. (We note that this construction
is equivalent to replacing $\ks$ by $\R\times\Sigma$ equipped with the
obviously defined metric). We can then conclude, as before,
that $(\hat \ks, \pi^*g)$ is isometrically diffeomorphic to the
Schwarzschild space--time. Hence $\ks$ is a quotient of
$(M^{\mathrm{Schw}},g^{\mathrm{Schw}})$ by a discrete subgroup $G$ of
the isometry group of $(M^{\mathrm{Schw}},g^{\mathrm{Schw}})$.

Now, one easily finds that in standard coordinates on the
Schwarzschild space--time every element $g$ of its isometry group acts
as follows:
\begin{equation}
  \label{eq:gact}
  {\R}\times(2m,\infty)\times S^2\ni (t,r,q)\longrightarrow g(t,r,q)=
  (\epsilon t+a,r,\omega(q))\ ,\qquad \epsilon = \pm 1\ ,
\end{equation}
where $\omega$ belongs to the isometry group of $S^2$ equipped with
the standard metric. Our hypothesis that $\Sext$ is diffeomorphic to
$\R^3$ minus a ball implies that $a\ne 0$ unless $\omega =
\mathrm{id}$. If there exists $g\in G$ for which $a\ne 0$, then
$\Mext$ will not have $\R\times \Sext$ topology.  This establishes the
second part of the Corollary under the hypothesis that $\Mext$ is
diffeomorphic to $\R\times \Sext$.

Let us finally show that if $G$ is not the trivial group, then there
exist closed timelike curves in $\Mext$. The hypothesis that $(M,g)$
is time orientable implies that $\epsilon=1$ in \eq{eq:gact}.  Let $
M^{\mathrm{Schw}}\ni p=(t,r,q)$ and consider the sequence
$g^n(p)=(t+na,r,\omega^n(q))$. It is easily seen that for $n$ large
enough we will have $g^n(p)\in I^+(p)$, thus there exists $n$ and a
future directed timelike curve $\gamma$ which starts at $p$ and ends
at $g^n(p)$. This holds for any $p\in M ^{\mathrm{Schw}}$, in
particular it will hold for $p$'s which are in $\Sext$. In that case
$\gamma$ can be chosen to lie in $\pi^{-1}(\Mext)$.  Then
$\pi(\gamma)$ will be a closed timelike curve in $\Mext$, and the
proof of the second part of the Corollary is complete. The first part
of the Corollary follows immediately from what has been said above.
\qed

To prove Theorem \ref{T1.1} we shall need the following minor
generalization of a result of Carter, which does not require
staticity:

\begin{Lemma}[Carter {\protect\cite[Corollary, p.~136]{CarterlesHouches}}]
  \label{lp1} Consider a space--time $(M,g)$ with a Killing vector
  field $X$ which is timelike in an asymptotically flat end $\Sext$,
  and suppose that the orbits of $X$ are complete in $\doc$. If there
  are no closed timelike curves through $\Sext$ contained in $\doc$,
  then the Killing vector field $X$ is nowhere vanishing on $\doc$.
\end{Lemma}

\proof Since $\doc$ is an open subset of $M$ invariant under $\phi_t$
we can without loss of generality assume that $M=\doc$. Suppose that
there exists $p\in\doc$ such that $X(p)=0$, then $p$ is invariant
under the flow of $X$, and so is $\partial J^+(p)$ because
$\phi_t(\partial J^+(p)) = \partial(\phi_t(J^+(p))) = \partial
J^+(\phi_t(p))$. Now since $p\in\doc$ there exists a future directed
timelike curve $\gamma$ which starts at $p_-\in\Mext$, passes through
$p$ and finishes at $p_+\in \Mext$.

From $J^+(p)\ni p_+\in \Mext$ we have $ J^+(p)\cap \Mext \ne
\emptyset$.
Now $\partial J^+(p)\cap \Mext$ is invariant under $\phi_t$ because
both $\partial J^+(p)$ and $\Mext$ are. The set $\partial J^+(p)\cap
\Mext$, if non--empty, is thus a null hypersurface in $\Mext$, and
since there are no $\phi_t$ invariant null hypersurfaces in $\Mext$ we
obtain $\partial J^+(p)\cap \Mext = \emptyset$. Hence $$J^+(p)\cap
\Mext = \Mext\ .$$
It follows that there exists a future directed
timelike curve $\gamma'$ from $p$ to $p^-$. Then the future directed
timelike curve $\gamma''$ obtained by following $\gamma$ from $p_-$ to
$p$ and then $\gamma'$ from $p$ to $p_-$ is a closed timelike curve
through $p_-$. As $p_-\in\Mext$ there exists $t\in\R$ such that
$p_-=\phi_t(q)$, for some $q\in\Sext$. Then $\phi_{-t}(\gamma'')$ is a
future directed timelike curve from $\Sext$ to itself, which gives a
contradiction, and the result follows.  \qed

\medskip

We are ready now to pass to the

\medskip

\noindent\textsc{Proof of Theorem \protect\ref{T1.1}:} Let $\dS$ be that
connected component of the set $\{q\in\Sigma\ |\ (X^\mu X_\mu)(q)<
0\}$ which contains the chosen asymptotic end $\Sext$.
$\overline{\dS}$ clearly has compact interior, therefore $\dS$
satisfies the hypotheses of Theorem \ref{T1}.  Moreover under all
alternative conditions, except perhaps for condition \cref{simplynew},
the hypotheses of Corollary \ref{c1} are satisfied with $\Sigma$
replaced by $\dS$. We can thus conclude, except perhaps in the case
\cref{simplynew}, that the Killing development $\kds$ of $\ds$ is
isometrically diffeomorphic to the d.o.c. of the
Schwarzschild--Kruszkal--Szekeres space--time. It follows from the
proof of Corollary \ref{c1} that in the case \cref{simplynew} the
Killing development $\kds$ of $\ds$ is isometrically diffeomorphic to
a quotient of the d.o.c. of the Schwarzschild--Kruszkal--Szekeres
space--time by an appropriate (perhaps trivial) subgroup of its
isometry group.

Next, let us observe that in a globally hyperbolic space--time any
d.o.c.  is globally hyperbolic, which is easily seen using the
definition of global hyperbolicity that involves compactness of the
sets of $J^+(p)\cap J^-(q)$. This shows that the alternative condition
\cref{Mgh} implies the alternative condition \cref{docgh}.

Further, a globally hyperbolic vacuum d.o.c. is simply connected by
\cite{ChWald}. Thus the alternative condition \cref{docgh} implies the
alternative condition \cref{simply}. That last alternative condition
is taken care of by the following:

\begin{Proposition}[Carter {\protect{\cite[Section 4]{CarterlesHouches}}}]
  \label{pcarter}
  Under the main hypotheses of Theorem \ref{T1.1}, suppose
  further that the alternative condition \cref{simply}
  holds.  Then the conclusion of Theorem \ref{T1.1} holds.
\end{Proposition}

\proof Let $\ddoc$ denote that connected component of the set
$\{q\in\doc\ |\ (X^\mu X_\mu)(q)< 0\}$ which contains the chosen
asymptotic end $\Sext$.  Clearly
$$\ddoc=\kds(=\cup_{t\in\R} \phi_t(\ds)),$$
where $\dS$ is that
connected component of the set $\{q\in\Sigma\ |\ (X^\mu X_\mu)(q)<
0\}$ which contains the chosen asymptotic end $\Sext$. Suppose that
$\kds\ne \doc$, then $\pd \kds \ne \emptyset$, where $\pd \Omega$
denotes the boundary of a set $\Omega$ in $\doc$.  By Lemma
\ref{lp1} $X$ has no zeros on $\pd \kds$, and Lemma \ref{lcvnew}
 \ptc{the argument here requires $\pd\kds$ to contain  a closed
 embedded hypersurface if non-empty; and it is not clear how to justify this
 without  further hypotheses; compare the Addendum}
shows that $\pd \kds$ is a Killing horizon. In particular $\pd \kds$
is a smooth null hypersurface with $\kds$ lying, locally, at one side
of it, so that there exists a smooth null vector field $\ell$ which is
transverse to $\pd\kds$ and which points outwards from $\kds$. (Such
vector fields can be defined locally by using local two--dimensional
cross--sections of the horizon, and requiring $\ell$ to be normal to
those cross--sections. Those locally defined vector fields can be
patched together to a globally defined one on any connected component
of $\pd \kds$ by using a partition of unity.) Now with our conventions
$X$ is everywhere future pointing on $\pd \kds$, but $\ell$ can be
future pointing at some points, and past pointing at some others, so
we set
$$
N_+=\{{p\in \pd \kds\ | \ g(X,\ell)< 0}\}\ ,\quad
N_-=\{{p\in \pd \kds\ | \ g(X,\ell)> 0}\}\ .
$$
While $N_+$ and $N_-$ do not have to be closed in $M$, Lemma
\ref{lcvnew} shows that $N_+$ and $N_-$ are closed in $\doc$.
Proposition \ref{Pkilling} shows that the set $\cN$ defined in the
proof of Lemma \ref{lcvnew} separates an appropriately small
neighborhood $\cV$ of $p$ into the local future of $\cN$ and its local
past, with $\kds\cap\cV$ lying to the local past of $\cN$:
\begin{equation}
  \label{eq:empty0}
  I^+(\pds;\cV)\cap \kds = \emptyset \ .
\end{equation} Here
$I^+(\Omega;N)$ denotes the chronological future of a set $\Omega$ in
a space--time $N$. In particular no future directed causal curve
through $\pds$ can enter $\kds$ through $N_+$. Similarly no future
directed causal curve through $\pds$ can leave $\kds$ through $N_-$.

Let $p\in N_+$; since $N_+\subset \doc$ there exists a timelike future
directed curve $\gamma$ from $p$ to $\Mext$. Now $\gamma$ leaves
$\kds$ at $p$ so clearly it has to reenter $\kds$ again through some
point $q\in N_-$. Let $q$ be the first such point, consider the path
$\cal P$ obtained by following $\gamma$ from $p$ until a little beyond
$q$ to $\kds$, and then following any curve contained entirely in
$\kds$ until $p$.  Since $N_+$ is closed the intersection number of
$\cal P$ and $ N_+$ is a well defined homotopy invariant (\emph{cf.,
  e.g.}, \cite[Chapter 3]{GuilleminPollack}\footnote{We are grateful
  to G.~Galloway for pointing out this reference and for a
  simplification of a previous version of this argument.}) and equals
one.  It follows that $\cal P$ cannot be deformed to a trivial loop
the image of which is a point, 
as such loops have vanishing intersection number with $ N_+$.  This
contradicts the hypothesis that $\doc$ is simply connected, and the
proposition follows.  \qed

The alternative condition \cref{simplynew} is taken care of by the
following:

\begin{Proposition}
  \label{psimplynew}
  Under the main hypotheses of Theorem \ref{T1.1}, suppose
  instead that the alternative condition \cref{simplynew}
  holds.  Then the conclusion of Theorem \ref{T1.1} holds.
\end{Proposition}

\proof The result follows by a repetition of the proof of Proposition
\ref{pcarter} as applied to the space--time $M'\equiv
M\setminus\{p|X(p)=0\}$, with the metric obtained by restriction of
$g$ to $M'$. We note that Lemma \ref{lp1} is not needed anymore as $X$
has no zeros in $M'$ by construction. The argument of the proof of
Proposition \ref{pcarter} shows that $\kds=\doc\setminus\{p|X(p)=0\}$.
If $\kds$ were a non--trivial quotient of the d.o.c. of the
Schwarzschild--Kruszkal--Szekeres space--time it would not be simply
connected. As $\doc\setminus\{p|X(p)=0\}$ is simply connected by
hypothesis, it follows that $\kds$ is the d.o.c. of the
Schwarzschild--Kruszkal--Szekeres space--time. Now $\doc$ is open,
$\kds$ is open, and for any non--trivial Killing vector field $X$ the
set $\{p|X(p)=0\}$ has no interior. Those remarks and elementary
topological considerations imply that $\doc\setminus\{p|X(p)=0\}=
\doc$,

It remains to consider
the alternative conditions \cref{nowhite}, \cref{nozeros} and
\cref{allzeros}. We note the following:

\begin{Lemma}
  \label{lnowhite} Under the main hypotheses of Theorem \ref{T1.1},
  suppose instead that the alternative condition \cref{nowhite}
    holds.  Then $X$ has no zeros on $\partial \Sigma$.
\end{Lemma}

\proof
Suppose that there exists $p\in\partial\Sigma$ such that
$X(p)=0$. The hypothesis that there is no white hole region implies
that $M=I^+(\Mext)$, hence $J^{-}(p)\cap \Mext\ne \emptyset$. As in
the proof of Lemma \ref{lp1} we conclude that $J^{-}(p)\supset
\Mext$. It follows that there exists a future directed timelike curve
$\gamma$ from a point $q\in I^{+}(\Sext)$ to $p\in \partial\Sigma$,
which contradicts achronality of $\overline\Sigma$; a contradiction
with achronality of $\Sigma$ easily follows.
\qed

Lemma \ref{lnowhite} reduces the alternative condition \cref{nowhite}
to the alternative condition \cref{nozeros}, which we consider now:

\begin{Proposition}
  \label{pnowhite} Under the main hypotheses of Theorem \ref{T1.1},
  suppose instead that the alternative condition \cref{nozeros} holds.
  Then the conclusion of Theorem \ref{T1.1} holds.
\end{Proposition}

\proof
Let, as before, $\dS$, denote that connected component of the set
$\{q\in\Sigma\ |\ (X^\mu X_\mu)(q)< 0\}$ which contains the chosen
asymptotic end $\Sext$.  By Lemma \ref{lp1} the Killing vector does
not vanish on $\Sigma$, and it does not vanish on $\partial \Sigma$ by
hypothesis, therefore $X$ has no zeros on ${\dS}\cup\pds \subset
\overline{\Sigma}$. By Theorem \ref{T1} $\pdS$ is connected and
has a well defined outwards pointing unit normal vector $n$
with respect to the induced metric.  Let $T$ denote the future
pointing unit normal to $\Sigma$, the non--vanishing of $X$ and
the connectedness of $\pdS$ imply that $X$ must be proportional
either to $T+n$ or $T-n$ on $\pdS$. Changing the time
orientation and $X$ to $-X$ if necessary we may without loss of
generality suppose that $X$ is proportional to $T-n$ on $\pdS$,
with a strictly positive proportionality factor. Proposition
\ref{Pkilling} shows that there exists a sufficiently small
neighborhood $\cV$ of $\pdS$ such that the set $\cN$ defined in
the proof of Lemma \ref{lcvnew}
 \ptc{the argument here requires $\pd\kds$ to contain  a closed
 embedded hypersurface if non-empty; and it is not clear how to justify this
 without further hypotheses}
separates $\cV$ into the local future of $\cN$ and its local
past, with $\kds\cap\cV$ lying to the local past of $\cN$:
\begin{equation}
  \label{eq:empty}
  I^+(\pds;\cV)\cap \kds = \emptyset\ .
\end{equation}
In particular every future directed causal curve through $\pds$ leaves
$\kds$ when crossing $\pds$.

Suppose that $\pdS\cap \doc\ne \emptyset$, thus there exists a point
$p\in \pdS$ and a timelike curve $\hg:[0,2]\to\doc$ such that
$\hg(0)=p_-\in\Mext$, $\hg(1)=p\in \pds$, and $\hg(2)=p_+\in\Mext$.
Define $\gamma:\R\to\doc$ by
$$
\gamma(s)=\begin{cases}{ \phi_s(p_-), }& s\in (-\infty,0], \\
  \hg(s),  & s\in [0,2],  \\
 \phi_{s-2}(p_+), & {s\in [2,\infty).}\end{cases}
$$
Then $\gamma$ is an inextendible timelike curve through $p\in\pds$,
with $\gamma(s)\in \Mext \subset \kds$ for $s\le 0$ and for $s\ge 2$,
and with $\gamma(1)=p\not\in \ds$.

Let $I_+$ be that connected component of $\gamma\cap\kds$ which
contains $[2,\infty]$.  Now $I_+\ne \R$ since $\gamma(1)=p$ and $p\not\in
\kds$; this last assertion follows from the fact that $X$ is timelike
throughout $\kds$ while $X$ is null at $p$. Since $\kds$ is open we
obtain $I_+=(s_+,\infty)$, for some $s_+\in (1,2)$. Set
$$
\tg= \gamma\Big|_{(s_+,\infty)}\ ;
$$
thus $\tg$ is an inextendible timelike curve in $\kds$. As $\kds$
is $\phi_t$ invariant, we have that for all $t$ the curve
$\phi_t(\gamma)$ is an inextendible timelike curve in $\kds$.

We claim that $\tg$ has to intersect $\ds$. To see that this must be
the case, define
$$
I=\{t\in \R: \phi_t(\tg)\cap\ds \ne \emptyset\}\ .
$$
Consider, first, the point $p_+\in \Mext$ defined above. By
definition of $\Mext$ we can write $p_+=\phi_{t_+}(q_+)$ for some
$q_+\in \Sext \subset \ds$. It follows that
$\phi_{-t_+}(\gamma)\ni\phi_{-t_+}(p_+)=q_+\in \ds$, so that $-t_+\in
I$. In fact by the definition of $\gamma$ we have that
$\phi_{-s}(\gamma)\cap \Sext=q_+\in \ds$ for all $s>t_+$, thus
$I\supset (-\infty,-t_+]$,
in particular $I\ne \emptyset$.

Note that $\phi_t(\gamma)$ is timelike for all $t$, hence transverse
to $\Sigma$ for all $t\in I$, which implies that $I$ is open. Let
$\hI$ be the connected component of $I$ containing $(-\infty,-t_+]$
and suppose that $\hI\ne\R$, then there exists $T_-\in \R$ such that
$\hI= (-\infty,T_-)$. Let $t_i\in\hI$ be any sequence converging to
$T_-$, set $q_i= 
\phi_{t_{i}}(\gamma)\cap \ds$. By interior compactness of
$\overline{\ds}$, passing to a subsequence if necessary, we may
suppose that $q_i$ converges to $q\in \overline{\ds}$. Clearly $q\in
\pds$ by the definition of $T_-$, hence $\phi_{T_-}(\gamma)$ is a
timelike curve through $q\in \pds$ which immediately enters $\kds$.
There are however no such curves by \eq{eq:empty}, hence $\hI=I=\R$.
In particular $0\in I$ so that $\tg\cap\ds=\phi_0(\tg)\cap\ds\ne
\emptyset$. Let $r$ be any point in $\tg\cap\ds$, we have $p\ne r$ as
$p\not \in \ds$. It follows that $\gamma$ is a timelike curve which
meets $\Sigma$ at two distinct points $p$ and $r$, which contradicts
achronality of $\Sigma$, and the Lemma  follows. \qed

It remains to show that Theorem \ref{T1.1} holds under the
alternative condition \cref{allzeros}. Suppose, thus, that there
exists a point $p\in 
\pds\cap \doc$ at which the Killing vector vanishes. By Boyer's
Theorem \ref{tboyer} $p$ belongs to a ``bifurcation surface'' of a
``bifurcate Killing horizon''. In particular there are four Killing
vector orbits which accumulate at $p$ and which coincide with two
geodesic generators of $\partial
J(p)=\partial (J^{+}(p)\cup J^{-}(p))$ with $p$ removed.
We note the following:

\begin{Lemma}
  \label{lnonempty}  Under the main hypotheses of Theorem \ref{T1.1},
  suppose that there exists a point $p\in 
  \pds\cap \doc$ at which the Killing vector vanishes. If $\Sigma$ is
  achronal, then the Killing orbits in $\partial J(p)\cap
  \partial\kds$ do \emph{not} belong to $\doc$.
\end{Lemma}

\proof The local structure of the orbits of the flow of the Killing
vector near $p$ shows that we can find a point $q\in \partial
J^+(p)\cap \partial\kds$ with the following properties:
\begin{enumerate}
\item The orbit of $X$ through $q$ coincides with the future directed
  null geodesic generator of the Killing horizon accumulating at $p$
  in the past.
\item $q\in I^+(\Sigma)\ .$
\end{enumerate}
We wish to show that for all $t\in\R$ we will have
\begin{equation}
  \label{eq:tinr}
  \phi_t(q)\in  I^+(\Sigma)\ .
\end{equation}
To establish \eq{eq:tinr}, let $\gamma$ be a timelike future directed
curve from $\Sigma$ to $q$, for $t\ge 0 $ consider the future directed
causal curve from $\Sigma$ to $q$ obtained by following $\gamma$ from
$\Sigma$ to $q$ and then following the orbit $\phi_s(q)$ of $X$ from
$q=\phi_0(q)$ to $\phi_t(q)$; the curve so constructed is not a null
geodesic and can therefore be deformed to a timelike curve.  On the
other hand, for $t\le 0$ the curve $\phi_t(\gamma)\cap J^+(\Sigma)\ne
\emptyset$ provides the desired timelike curve from $\Sigma$ to
$\phi_t(q)$.

Suppose that $q\in\doc$, 
then there exists a future directed timelike curve $\hat \gamma$ from
$q$ to a point $p_+\in \Mext$. By definition of $\Mext$ we have
$p_+=\phi_{t_+}(q_+)$ for some $q_+\in \Sext$ and some $t_+\in \R$.
Then $\phi_{-t_+}(\hat \gamma)$ is a future directed timelike curve
passing through $\phi_{-t_+}(q)\in I^+{(
\Sigma)}$ and $q_+\in \Sext$,
which contradicts achronality of $\Sigma$. This establishes the result
for $\partial J^{+}(p)\cap\partial \kds$. The result for $\partial
J^{-}(p)\cap\partial \kds$ follows from this one by changing the time
orientation of $(M,g)$.  \qed

Lemma \ref{lnonempty} immediately implies Theorem \ref{T1.1} under the
alternative condition \cref{allzeros}: indeed, under this condition
the Killing vector field $X$ vanishes throughout $\pds$. It follows
that the set of those zeros of the Killing vector field which lie on
$\pkds$ is a compact connected manifold.  It is then easily seen that
each Killing orbit on $\pkds$ is of the form $\partial
J^{+}(p)\cap\partial \kds$ or $\partial J^{-}(p)\cap\partial \kds$ for
some $p\in\partial\dS$, and the proof of Theorem \ref{T1.1} is
completed.  \qed

\section{Concluding remarks}
\label{conclusions}

In this paper we have essentially finished the classification of
asymptotically flat, static, vacuum, appropriately regular black hole
space--times.  It is natural to try to generalize our results to
electro--vacuum space--times.  Recall that Simon \cite{Simon:elvac}
and Masood--ul--Alam \cite{Masood} have shown\footnote{The paper by
  Ruback \cite{Ruback} with similar claims contains essential gaps.},
roughly speaking, that all appropriately regular such black holes which
\emph{do not contain degenerate horizons} belong to the
Reissner--Nordstr\"om family. In the case of a \emph{connected} black
hole the requirement of non--degeneracy of the event horizon has been
removed by Heusler \cite{heuslerMP} (\emph{cf.\/} also
\cite{ChAscona,ChNad}); in \cite{heuslerMP} some partial results
concerning the case where \emph{all horizons are degenerate} have also
been obtained.  Nevertheless the general case of a static electro--vacuum
black hole containing both degenerate and non--degenerate horizons
remains open.  It turns out that the arguments used here can be
generalized to exclude some further classes of electro--vacuum black
holes without, so far, leading to a definitive classification of
electro--vacuum static black holes. We hope to be able to improve those
results in the future.

Let us close this paper with some comments concerning the
classification problem of stationary black holes. Recall that the
usual approach to the classification of \emph{ analytic}
electro--vacuum black holes is via the so--called \emph{Hawking
  rigidity theorem}\footnote{This theorem is wrong as stated in
  \cite{HE}, \emph{cf.\/} \cite{Ch:rigidity}. A corrected version can
  be found in \cite{ChAscona}.} which shows, under appropriate
hypotheses including analyticity, that event horizons in stationary
space--times with defining Killing vector $X$ are Killing horizons for
an appropriately defined Killing vector field $Y$. If $X=Y$ and if all
horizons are \emph{non--degenerate} it follows from the results in
\cite{Sudarsky:wald,ChWald1,RaczWald2}
\ptc{a simple proof can be found in~\cite[Section~7.2]{ChCo}.}
that the space--time must be static, so that the Israel
--- Bunting --- Masood--ul--Alam theorem (or its extension
here) can be used to analyze this case.  The case with $X=Y$
and \emph{all} horizons degenerate has been excluded in Section
\ref{Komar}. The case $X=Y$ and \emph{some} horizons degenerate
is still open, and it would be of interest to fill this gap.

Concerning the case $X\ne Y$, a complete classification \emph{modulo
  existence of ``struts'' on the symmetry axis} has been given by
Weinstein \cite{Weinstein3}, again assuming \emph{non--degeneracy} of
all event horizons.
\ptc{see~\cite{ChCo} for a detailed exposition, filling in the
gaps of the previous arguments}
We note that
even the question, whether a degenerate, connected, vacuum,
stationary--axi--symmetric, regular black hole is an extreme
Kerr black--hole has not been resolved so far.  Our analysis in
Section \ref{SBc} provides, we believe, a good starting point
for an extension of the results of \cite{Weinstein3} which
allows degenerate components of the event horizon.

\medskip

\appendix
\section{Addendum, June 2010}
\newcommand{\Spn}{S^+_0}
\newcommand{\Spz}{S^+_0}
\newcommand{\mcEp}{{\mcE^+}}
\newcommand{\mcEpz}{{\mcE^+_0}}
\newcommand{\mcEm}{{\mcE^-}}
\newcommand{\mcEmz}{{\mcE^-_0}}

\newcommand{\hypot}{\,\,\mathring{\!\! \hyp_t}}
\newcommand{\hypoz}{\,\,\mathring{\!\! \hyp_0}}
\newcommand{\ohypoz}{\overline{\hypoz}}
\newcommand{\ohypot}{\overline{\hypot}}

\newcommand{\odoc}{\overline{\doc}}
\newcommand{\ohyp}{\,\,\overline{\!\!\hyp}}
\newcommand{\pohyp}{\partial\ohyp}

\newcommand{\aregular}{{an {\regular}}}
\newcommand{\regular}{$I^+$--regular}

\newcommand{\hthreeg}{h}

\newcommand{\doce}{\doc_\epsilon}

\newcommand{\odocIp}{\overline{\doc}\cap I^+(\Mext)}
\newcommand{\odocup}{{\doc}\cup \mcH_0^+}
\newcommand{\llambda}{\lambda}

\newcommand{\ue}{u_{\epsilon}}
\newcommand{\uee}{u_{\epsilon,\eta}}

\newcommand{\bmcM}{\,\,\,\,\widetilde{\!\!\!\!\mcM}}
\newcommand{\bfourg}{\widetilde{\fourg}}

\newcommand{\eean}{\nonumber\end{eqnarray}}

\newcommand{\lp}{\ell}

\newcommand{\tSp}{\tilde S_p}
\newcommand{\Sp}{S_p}

\newcommand{\dgtcp}{\dgt}
\newcommand{\dgtc}{\dgt}

\newcommand{\id}{{\rm id}}

\newcommand{\Zd}{{\mcZ}_{\dgt}}

\newcommand{\mcHp}{{\mcH^+}}
\newcommand{\mcHpz}{{\mcH^+_0}}
\newcommand{\mcHm}{{\mcH^-}}
\newcommand{\mcHmz}{{\mcH^-_0}}

\newcommand{\zN}{\mathring N}

\newcommand{\mcY}{{\mycal Y}}
\newcommand{\mcX}{{\mycal X}}

\newcommand{\Nndg}{N_{\mbox{\scriptsize\rm  ndh}}}
\newcommand{\Ndg}{N_{\mbox{\scriptsize\rm  dh}}}
\newcommand{\Nndh}{N_{\mbox{\scriptsize\rm  ndh}}}
\newcommand{\Ndh}{N_{\mbox{\scriptsize\rm  dh}}}
\newcommand{\Naf}{N_{\mbox{\scriptsize\rm  AF}}}
\newcommand{\alpharate}{\lambda}
\newcommand{\zalpha}{\mathring \alpha}

\newcommand{\puncti}{a_i}
\newcommand{\ai}{\puncti}

\newcommand{\mcS}{{\mycal S}}

\newcommand{\nic}{}
\newcommand{\phypa}{\partial\mcS_{(a)} }
\newcommand{\Ya}{{}^{(a)}Y }
\newcommand{\Ja}{{}^{(a)}J }
\newcommand{\Oma}{{}^{(a)}\Omega }
\newcommand{\Xa}{{}^{(a)}X }
\newcommand{\Sa}{{}^{(a)}S }
\newcommand{\Qea}{{}^{(a)}Q^E}
\newcommand{\Qba}{{}^{(a)}Q^B}
\newcommand{\phiea}{{}^{(a)}\Phi^E}
\newcommand{\phiba}{{}^{(a)}\Phi^B}
\newcommand{\mcHa}{{}^{(a)}\mcH}

\newcommand{\hypo}{\mathring \hyp}
\newcommand{\ohypo}{\overline{\mathring \hyp}}

\newcommand{\Kz}{K_\kl 0}

\newcommand{\fourge}{\fourg_\epsilon}

\newcommand{\Cp}{ \mcC^+ }
\newcommand{\Cpe}{ \mcC^+ _\epsilon}
\newcommand{\Ct}{ \mcC^+_t}
\newcommand{\Ctm}{ \mcC^+_{t_-}}
\newcommand{\hStmR}{\hat S_{t_-,R}}
\newcommand{\hStR}{\hat S_{t,R}}
\newcommand{\hSzR}{\hat S_{0,R}}
\newcommand{\hSts}{\hat S_{\tau,\sigma}}

\newcommand{\Sone}{\Sz}
\newcommand{\Sonep}{\Sz}
\newcommand{\Soneq}{S_{0,q}}
\newcommand{\Stp}{S_{t,p}}
\newcommand{\Stq}{S_{t,q}}
\newcommand{\Sz}{S_0}

\newcommand{\bw}{\bar w}
\newcommand{\bzeta}{\bar \zeta}

\newcommand{\zMtwo}{\mathring{M}{}^2}
\newcommand{\Mtwo}{{M}{}^2}
\newcommand{\bMtwo}{{\bar M}{}^2}
\newcommand{\hMtwo}{{\hat M}{}^2}

\newcommand{\hypext}{\hyp_{\mbox{\scriptsize \rm ext}}}
\newcommand{\Mtext}{\Sext}
\newcommand{\Mint}{\mcM_{\mbox{\scriptsize \rm int}}}
\newcommand{\mcMext}{\Mext}

\newcommand{\mcHN}{\mcN}
\newcommand{\mcNH}{\mcH}

\newcommand{\hSq}{{\hat S_q}}
\newcommand{\hSp}{{\hat S_p}}
\newcommand{\hSpn}{{\hat S_{p_n}}}
\newcommand{\hSqn}{{\hat S_{q_n}}}
\newcommand{\mcA}{\mycal A}
\newcommand{\mcZ}{\mycal Z}

\newcommand{\zh}{{\,\,\widetilde{\!\!\mcZ}}}
\newcommand{\dgt}{{\mycal Z}_{dgt}}

\newcommand{\kl}[1]{{(#1)}}

\newcommand{\Uone}{{\mathrm{U(1)}}}

\newcommand{\Sm}{\ensuremath{\Sigma_{-}}}
\newcommand{\Nt}{\ensuremath{N_{2}}}
\newcommand{\Nth}{\ensuremath{N_{3}}}

\newcommand{\jlca}[1]{\mnote{{\bf jlca:} #1}}

\newcommand{\jlcared}[1]{{\color{red}\mnote{{\color{red}{\bf jlca:}
#1} }}}
\newcommand{\jlcachange}[1]{{\color{blue}\mnote{\color{blue}{\bf jlca:}
changed on 18.IX.07}}{\color{blue}{#1}}\color{black}\ }

\newcommand{\jlcachangeb}[1]{{\color{blue}\mnote{\color{blue}{\bf jlca:}
changed on 19.IX.07}}{\color{blue}{#1}}\color{black}\ }
\newcommand{\jlcachangec}[1]{{\color{blue}\mnote{\color{blue}{\bf jlca:}
changed on 20.IX.07}}{\color{blue}{#1}}\color{black}\ }
\newcommand{\jlcachanged}[1]{{\color{blue}\mnote{\color{blue}{\bf jlca:}
changed after 9.X.07}}{\color{blue}{#1}}\color{black}\ }
\newcommand{\jlcachangeXIII}[1]{{\color{blue}\mnote{\color{blue}{\bf jlca:}
changed after 13.X.07}}{\color{blue}{#1}}\color{black}\ }

\newcommand{\jlcaadded}[1]{{\color{blue}\mnote{\color{blue}{\bf jlca:}
added on 20.IX.07}}{\color{blue}{#1}}\color{black}\ }
\newcommand{\sse }{\Longleftrightarrow}
\newcommand{\se}{\Rightarrow}
\newcommand{\uu}{{\bf u}}
\newcommand{\bbN}{\mathbb{N}}
\newcommand{\bbZ}{\mathbb{Z}}
\newcommand{\bbC}{\mathbb{C}}
\newcommand{\bbH}{\mathbb{H}}

\newcommand{\cof}{\operatorname{cof}}
\newcommand{\dive}{\operatorname{div}}
\newcommand{\curl}{\operatorname{curl}}
\newcommand{\grad}{\operatorname{grad}}
\newcommand{\rank}{\operatorname{rank}}
\def\G{{\mycal G}}
\def\scro{{\mycal O}}
\def\Doc{\doc}
\def\scrip{\scri^{+}}%
\def\scrp{{\mycal I}^{+}}%
\def\Scri{\scri}
\def\scra{{\mycal A}}
\def\Scra{\cup_{i}\scra_i}
\def\scru{{\mycal U}}
\def\scrw{{\mycal W}}
\def\scrv{{\mycal V}}
\def\scrs{{\mycal S}}
\def\rot{{\mycal R}}
\def\khor{{\mycal H}}
\def\e{\wedge}
\def\d{\partial}
\def\KK{\phi^K}
\def\K0{\phi^{K_0}}
\def\Kdot{\phi^{K}\cdot}
\def\X.{\phi^{X}\cdot}
\def\normK{W}

\newcommand{\hyphat}{\,\,\,\widehat{\!\!\!\hyp}}

\newcommand{\newF}{\lambda}

\newcommand{\Int}{\operatorname{Int}}
\newcommand{\Tr}{\operatorname{Tr}}

\newcommand{\myuu}{u}

\newcommand{\oX}{\overline X}
\newcommand{\oY}{\overline Y}
\newcommand{\op}{\overline p}
\newcommand{\oq}{\overline q}

\newcommand{\mcEh}{{{\mycal E}^+_\hyp}}

{\catcode `\@=11 \global\let\AddToReset=\@addtoreset}
\AddToReset{equation}{section}
\renewcommand{\theequation}{\thesection.\arabic{equation}}

\newcommand{\ptcKcite}[1]{.\cite{#1}}

\newcommand{\fourg}{{\mathfrak g }}

\newcommand{\refcite}[1]{\cite{#1}}
\newcommand{\levoca}[1]{}

\newcommand{\umac}{\gamma}%
\newcommand{\metrict}{\threeg }%

\newcommand{\mcN}{{\mycal N}}
\newcommand{\mcNX}{{\mycal N(X)}}

\newcommand{\cf}{cf.,}

\newcommand{\gcirc}{{\mathring{g}}_{ab}}
\let\a=\alpha\let\b=\beta \let\g=\gamma \let\d=\delta\let\lb=\lambda
\newcommand{\sts}{spacetimes}
\newcommand{\calM}{{\mcM}}
\newcommand{\calH}{{\mcH}}
\newcommand{\el}{e^{2\lb}}
\newcommand{\hysf}{hypersurface}
\newcommand{\cH}{{\mycal  H}}
\newcommand{\OOmega}{\omega}

\newcommand{\nopcite}[1]{}

\newcommand{\hypM}{\hyp}

\newcommand{\gschw}{h_{\mbox{\scriptsize \rm Schw}}}
\newcommand{\gsch}{\gschw}

\newcommand{\calSJr}{{\mycal  S}_{(J_0,\rho_0)}}
\newcommand{\Ima}{\mbox{\rm Im}}
\newcommand{\Ker}{\mbox{\rm Ker}}
\newcommand{\sgstatic}{{strictly globally static{}}}
\newcommand{\gstatic}{{globally static{}}}
\newcommand{\riemg}{g}
\newcommand{\riemgz}{g_0}
\newcommand{\hhat}{\ghat}
\newcommand{\DeltaL}{\Delta_{\mathrm L}}
\newcommand{\ghat}{\gamma}
\newcommand{\hzhat}{\gamma_0}
\newcommand{\holder}{H\"older }
\newcommand{\del}{\partial}
\newcommand{\Mbar}{{\overline M}}
\newcommand{\gbar}{{\overline \riemg}}
\newcommand{\dm}{{\partial M}}
\newcommand{\dminfty}{{\partial_\infty M}}
\newcommand{\dmo}{{\partial_0 M}}
  {\renewcommand{\theenumi}{\roman{enumi}}

\newcommand{\bS}{{\overline \Sigma}}
\newcommand{\tp}{\tau_p}
\newcommand{\hF}{\hat F}
\newcommand{\signX}{\mathrm{sign}X}
\newcommand{\mnu}{\nu}
\newcommand{\homega}{{\widehat{\Omega}}}%
\newcommand{\kerp}{\mathrm{Ker}_p\nabla X}%
\newcommand{\kerpi}{\mathrm{Inv}_p}%
\newcommand{\const}{\mathrm{const}}
\newcommand{\ml}{{M_{0,n-2\ell}}}
\newcommand{\mil}{{M_{\mathrm{iso},\ell}}}
\newcommand{\mtwo}{{M_{0,n-2}}}
\newcommand{\mfour}{{M_{0,n-4}}}
\newcommand{\sell}{\pi_\Sigma(\ml)}
\newcommand{\sfour}{\pi_\Sigma(\mfour)}
\newcommand{\stwo}{\pi_\Sigma(\mtwo)}
\newcommand{\mi}{{M_{0,n-2i}}}
\newcommand{\mj}{{M_{0,n-2j}}}
\newcommand{\htau}{{\hat \tau}}
\newcommand{\ttau}{{\tilde \tau}}
\newcommand{\zM}{{\mathring{M}}}
\newcommand{\ztau}{{\mathring{\tau}}}
\newcommand{\zSigma}{{\mathring{\Sigma}}}
\newcommand{\sings}{{\Sigma_{\mathrm{sing,iso}}}}
\newcommand{\miso}{{M_{\mathrm{iso}}}}
\newcommand{\psings}{{\partial\Sigma_{\mathrm{sing,iso}}}}
\newcommand{\sing}{{\Sigma_{\mathrm{sing}}}}
\newcommand{\singt}{{\Sigma_{\mathrm{sing},0}}}
\newcommand{\psingt}{{\partial\Sigma_{\mathrm{sing},0}}}
\newcommand{\Ein}{\operatorname{Ein}}
\newcommand{\hlambda}{\hat \lambda}
\newcommand{\mLX}{{\mcL_X}}
\newcommand{\mcE}{{\mycal E}}
\newcommand{\mcC}{{\mycal C}}
\newcommand{\mcD}{{\mycal D}}
\newcommand{\mcW}{{\mycal W}}
\newcommand{\lormet  }{{\frak g}}
\newcommand{\ApSwSw}{Appendix~\ref{SwSs}}
\newcommand{\abs}[1]{\left\vert#1\right\vert}
\newcommand{\norm}[1]{\left\Vert#1\right\Vert}
\newcommand{\M}{\EuScript M}
\newcommand{\Lie}{\EuScript L}
\newcommand{\nablash}{\nabla{\kern -.75 em
     \raise 1.5 true pt\hbox{{\bf/}}}\kern +.1 em}
\newcommand{\Deltash}{\Delta{\kern -.69 em
     \raise .2 true pt\hbox{{\bf/}}}\kern +.1 em}
\newcommand{\Rslash}{R{\kern -.60 em
     \raise 1.5 true pt\hbox{{\bf/}}}\kern +.1 em}
\newcommand{\Div}{\operatorname{div}}
\newcommand{\dist}{\operatorname{dist}}
\newcommand{\Rfour}{\bar R}

\newcommand{\tthreeg}{\tilde\threeg}
\newcommand{\tcalD}{\widetilde\calD}
\newcommand{\tnabla}{\tilde \nabla}

\newcommand{\tD}{\tilde D}

\newcommand{\gammab}{\bar\gamma}
\newcommand{\chib}{\bar\chi}
\newcommand{\gb}{\bar g}
\newcommand{\Nb}{\bar N}
\newcommand{\Hb}{\bar H}
\newcommand{\Ab}{\bar A}
\newcommand{\Bb}{\bar B}
\newcommand{\betah}{{\hat\beta}}
\newcommand{\chit}{\tilde\chi}
\newcommand{\Ht}{\tilde H}
\newcommand{\Ric}{\operatorname{Ric}}
\newcommand{\supp}{\operatorname{supp}}
\newcommand{\bA}{\mathbf{A}}
\newcommand{\st}{\colon\>}

\newcommand{\hbound}{\mathring{\mathsf{H}}}
\newcommand{\cbord}{{\mathsf{C}}}
\newcommand{\pM}{\partial M}
\newcommand{\hM}{\widehat{M}}
\newcommand{\chyp}{\mathcal C}
\newcommand{\hhyp}{\mathring{\mathcal H}}
\newcommand{\bnabla}{\overline{\nabla}}
\newcommand{\maclK}{{\mathcal K}}
\newcommand{\maclKzo}{{\mathcal K}^{\bot_g}_0}
\newcommand{\maclKz}{{\mathcal K}_0}
\newcommand{\hbord}{\hbound}%
\newcommand{\cKi}{{\mycal K}_{i0}}
\newcommand{\mcO}{{\mycal O}}
\newcommand{\mcT}{{\mycal T}}
\newcommand{\mcU}{{\mycal U}}
\newcommand{\mcV}{{\mycal V}}
\newcommand{\eug}{{\frak G}}

\newcommand{\rd}{\,{ d}} 

\newcommand{\ourU}{\mathbb U}
\newcommand{\ourW}{\mathbb W}
\newcommand{\hyp}{{\mycal S}}

\newcommand{\bhyp}{\overline{\hyp}}

\newcommand{\bg}{{\overline{g}_\Sigma}}

\newcommand{\Bgamma}{{B}} 
\newcommand{\bmetric}{{b}} 

\newcommand{\Kp}{{\mathfrak g}} 
\newcommand{\KA}{p} 
\newcommand{\gthreeup}{\,{}^3g} 
\newcommand{\arcsh}{{\rm argsh}}

\newcommand{\threeg}{\gamma}

\newcommand{\detthreeg}{{\det(\threeg_{ij})}} 
\newcommand{\lapse}{{\nu}} 
\newcommand{\shift}{{\beta}} 
\newcommand{\threeP}{{\pi}} 
\newcommand{\tildetg}{{\tilde \threeg}} 
\newcommand{\hst}{{\breve{h}}} 

\newcommand{\zn} {\mathring{\nabla}} 
\newcommand{\znabla} {\mathring{\nabla}} 
\newcommand{\mn} {M} 

\newcommand{\eg}{e.g.,}

\newcommand{\can}{{\mathrm \scriptsize can}}

\newcommand{\mcM}{{\mycal M}}
\newcommand{\mcR}{{\mycal R}}
\newcommand{\mcH}{{\mycal H}}
\newcommand{\mcK}{{\mycal K}}
\newcommand{\notreP}{{\widehat P}}
\newcommand{\ninfty}{N_\infty}
\newcommand{\bea}{\begin{eqnarray}}
\newcommand{\beaa}{\begin{eqnarray*}}
\newcommand{\bean}{\begin{eqnarray}\nonumber}
\newcommand{\kidap}{KID-hole}
\newcommand{\kidhor}{KID-horizon}
\newcommand{\kidaps}{KID-holes}
\newcommand{\bel}[1]{\begin{equation}\label{#1}}
\newcommand{\beal}[1]{\begin{eqnarray}\label{#1}}
\newcommand{\beadl}[1]{\begin{deqarr}\label{#1}}
\newcommand{\eeadl}[1]{\arrlabel{#1}\end{deqarr}}
\newcommand{\eeal}[1]{\label{#1}\end{eqnarray}}
\newcommand{\eead}[1]{\end{deqarr}}
\newcommand{\eea}{\end{eqnarray}}
\newcommand{\eeaa}{\end{eqnarray*}}
\newcommand{\nn}{\nonumber}
\newcommand{\Hess}{\mathrm{Hess}\,}
\newcommand{\Ricc}{\mathrm{Ric}\,}
\newcommand{\Riccg}{\Ricc(g)}
\newcommand{\Lpsi}{L^{2}_{\psi}}
\newcommand{\Lpsione}{\zH1_{\phi,\psi}}
\newcommand{\Lpsitwo}{\zH^{2}_{\phi,\psi}}

\newcommand{\Lpsig}{L^{2}_{\psi}(g)}
\newcommand{\Lpsioneg}{\zH1_{\phi,\psi}(g)}
\newcommand{\Lpsitwog}{\zH^{2}_{\phi,\psi}(g)}
\newcommand{\Lpsikg}[2]{\zH^{#1}_{\phi,\psi}(#2)}

\newcommand{\divr }{\mbox{\rm div}\,}
\newcommand{\tr}{\mbox{\rm tr}\,}
\newcommand{\J}{\delta J}
\newcommand{\source}{\delta \rho}
\newcommand{\Eq}[1]{Equation~(\ref{#1})}
\newcommand{\Eqsone}[1]{Equations~(\ref{#1})}
\newcommand{\Eqs}[2]{Equations~(\ref{#1})-\eq{#2}}
\newcommand{\Sect}[1]{Section~\ref{#1}}
\newcommand{\Lem}[1]{Lemma~\ref{#1}}

\newcommand{\zmcH }{\,\,\,\,\mathring{\!\!\!\!\mycal H}{}}

%
%
%
%
%



\def \Reel{\mathbb{R}}
\def \C{\mathbb{C}}
\def \R {\Reel}
\def \Hyp{\mathbb{H}}
\newcommand{\mcL}{{\mycal L}}
\def \Nat{\mathbb{N}}
\def \Z{\mathbb{Z}}
\def \N {\Nat}
\def \Sphere{\mathbb{S}}

\newcommand{\bM}{\,\overline{\!M}}

\newcommand{\zHkpp}{\zHk_{\phi,\psi}}
\newcommand{\Hkpp}{H^k_{\phi,\psi}}
\newcommand{\zHk}{\zH^k}
\newcommand{\zH}{\mathring{H}}


\newcommand{\ednote}[1]{}

\newcommand{\mcmg}{$(\mcM,\fourg)$}


\newcommand{\cem}[1]{\textcolor{bluem}{\emph{ #1}}}
\newcommand{\cbf}[1]{\textcolor{bluem}{\bf #1}}
\newcommand{\rbf}[1]{\textcolor{red}{\bf #1}}

\newcommand{\chX}{\changedX}
\newcommand{\mcMX}{\mcM_X}

\newcommand{\xflat}{X^\flat}
\newcommand{\kflat}{K^\flat}
\newcommand{\yflat}{Y^\flat}
\newcommand{\Tau}{\tau}

\newcommand{\bcM}{{\,\,\,\overline{\!\!\!\mcM}}}
\newcommand{\tcM}{\,\,\,\,\widetilde{\!\!\!\!\cM}}
\newcommand{\hJ}{{\hat J}}
\newcommand{\sthd}{{}^{\star_g}}
\newcommand{\hG}{\widehat \Gamma}
\newcommand{\hD}{\widehat D}
\newcommand{\hxi}{\hat \xi}
\newcommand{\hxib}{\hat{\xib}}
\newcommand{\heta}{\hat \eta}
\newcommand{\hetab}{\hat{\etab}}
\newcommand{\home}{\hat{\omega}}
\newcommand{\homb}{\hat{\underline\omega}}

\newcommand{\cimkd}{{\mycal C}^\flat}
\newcommand{\cimku}{{\mycal C}^\sharp}
\newcommand{\proj}{\textrm{pr}}
\newcommand{\Mclosed}{\bcM}
\newcommand{\cg}{\,{\tilde {\!g}}}

\newcommand{\cM}{\mycal M}
\newcommand{\cE}{\mycal E}

\newcommand{\cAp}{{\mycal A}_{\mbox{\scriptsize phg}}}
\newcommand{\cApM}{\cAp(M)}
\newcommand{\stsg}{{\mathfrak g}}
\newcommand{\tf}{\widetilde f}
\newcommand{\complementaire}{\complement}
\newcommand{\remarks}{{\bf Remarks : }}

\newcommand{\mcB}{{\mycal B}}
\newcommand{\cL}{{\mycal  L}}
\newcommand{\decal}{{\mycal D}}
\newcommand{\cC}{{\mycal C}}
\newcommand{\cG}{{\mycal G}}
\newcommand{\cCak}{{\cC}^\alpha_k}
\newcommand{\backmg}{b} 
\newcommand{\Id}{\mbox{\rm Id}} 
\newcommand{\sconst}{\mbox{\scriptsize\rm const}} 

\newcommand{\chnindex}[1]{\index{#1}}
\newcommand{\chindex}[1]{\index{#1}}
\newcommand{\bhindex}[1]{\chindex{black holes!#1}}

\newcommand{\tA}{\theta_\Al}
\newcommand{\Hau}{{\mathfrak H}}
\newcommand{\Ar}{\mbox{\rm Area}}
\newcommand{\Arm}{\mbox{\rm Area}_{S_2}}
\newcommand{\Al}{{\mathcal Al}}
\newcommand{\Hausone}{\Hau1}
\newcommand{\Htwohone}{\Hau^{2}}
\newcommand{\calS}{{\mathscr S}}
\newcommand{\BA}{B_\Al}
\newcommand{\Leb}{{\mathfrak L}}

\newcommand{\eqs}[2]{\eq{#1}-\eq{#2}}

\newcommand{\cW}{{\mycal W}}

\newcommand{\cA}{{\mycal A}}

\def\nen{\nonumber}
\def\Tau {{\mycal  T}}

\newcommand{\pihyp}{\partial_\infty\hyp}
\newcommand{\cMext}{{\mcM}_{\ext}}
\newcommand{\timkd}{{\mycal T}^\flat}
\newcommand{\timku}{{\mycal T}^\sharp}

\newcommand{\ro}{\rho}
\newcommand{\roo}{\bar{\ro}}

\newcommand{\les}{\lesssim}
\newcommand{\ges}{\gtrsim}
\def\piT{{\,^{(\T)}\pi}}
\def\Qr{\mbox{Qr}}
\def\flux{{\mbox{Flux}}}
\def\pa{\partial}
\def\onab{\overline{\nabla}}
\def\Db{{\bf {\bar {\,D}}}}
\def\Dh{{\bf {\hat {\,{D}}}}}
\def\laph{\hat{\Delta}}
\def\Null{\dot{\NN}^{-}}
\def\EE{{\mycal  E}}
\def\GG{{\mycal  G}}
\def\OO{{\mycal  O}}
\def\QQ{{\mycal  Q}}
\def\UU{{\mycal  U}}
\def\VV{{\mycal  V}}
\def\TT{{\mycal  T}}
\def\exp{\,\mbox{exp}}
\def\P{{\bf P}}
\def\vphi{\varphi}

\def \endprf{\hfill  {\vrule height6pt width6pt depth0pt}\medskip}
\def\emph#1{{\it #1}}
\def\textbf#1{{\bf #1}}
\newcommand{\Us}{\mbox{$U\mkern-13mu /$\,}}
\newcommand{\eql}{\eqlabel}
\def\medn{\medskip\noindent}

\def\Db{{^{(\Lb)}  D}}

\def\a{{\alpha}}
\def\b{{\beta}}
\def\ga{\gamma}
\def\Ga{\Gamma}
\def\eps{\epsilon}
\def\La{\Lambda}
\def\si{\sigma}
\def\Si{\Sigma}
\def\ro{\rho}
\def\th{\theta}
\def\ze{\zeta}
\def\nab{\nabla}
\def\varep{\varepsilon}

\def\aa{{\underline{\a}}}
\def\bb{{\underline{\b}}}
\def\bb{{\underline{\b}}}
\def\omb{{\underline{\om}}}
\def\Lb{{\underline{L}}}
\def\aaa{{\bold a}}

\newcommand{\trchb}{\tr \chib}
\newcommand{\chih}{\hat{\chi}}
\newcommand{\xib}{\underline{\xi}}
\newcommand{\etab}{\underline{\eta}}
\newcommand{\chibh}{\underline{\hat{\chi}}\,}
\def\chih{\hat{\chi}}
\def\trch{\mbox{tr}\chi}
\def\tr{\mbox{tr}}
\def\Tr{\mbox{Tr}}
\def\Xb{{\underline X}}
\def\Yb{{\underline Y}}

\def\rhoc{\check{\rho}}
\def\sic{\check{\si}}
\def\bboc{\check{\bb}}
\def\muc{\tilde{\mu}}

\def\AA{{\mycal  A}}
\def\BB{{\mycal  B}}
\def\MM{{\mycal  M}}
\def\NN{{\mycal  N}}
\def\II{{\mycal  I}}
\def\FF{{\mycal  F}}
\def\HH{{\mycal  H}}
\def\JJ{{\mycal  J}}
\def\KK{{\mycal  K}}
\def\Lie{{\mycal  L}}
\def\DD{{\mycal  D}}
\def\PP{{\mycal  P}}
\def\HH{{\mycal  H}}

\def\LL{{\mathcal L}}

\def\A{{\bf A}}
\def\B{{\bf B}}
\def\F{{\bf F}}
\def\I{{\bf I}}
\def\J{{\bf J}}
\def\M{{\bf M}}
\def\L{{\bf L}}
\def\O{{\bf O}}
\def\Q{{\bf Q}}
\def\R{{\mathbb R}}
\def\U{{\bf U}}
\def\S{{\bf S}}
\def\K{{\bf K}}
\def\g{{\bf g}}
\def\t{{\bf t}}
\def\u{{\bf u}}
\def\p{{\bf p}}
\def\LLb{{\bf \Lb}}

\def\Ab{\underline{A}}
\def\Eb{\underline{E}}
\def\Bb{\underline{B}}
\def\Hb{\underline{H}}
\def\Rb{\underline{R}}
\def\Vb{\underline{V}}
\def\AAA{{\Bbb A}}
\def\RRR{{\Bbb R}}
\def\MMM{{\Bbb M}}
\def\T{{\Bbb T}}
\def\To{\T}

\def\Aoc{\check{A}}
\def\Roc{\check{R}}
\def\nabb{\overline{\nab}}
\newcommand{\nabbb}{\mbox{$\nabla \mkern-13mu /$\,}}
\def\Ml{{\mycal  M}_{\ell}}
\def\intws{{\,\,\,\int_0^s \mkern-30mu \om}\,\,\,\,}
\def\expp{\mbox{exp}}

\def\B{{\bf B}}

\newcommand{\changedX}{K}

\newcommand{\hahyp}{\,\,\widehat{\!\!\hyp}}%

\def\lap{\Delta}
\def\pr{\partial}
\newcommand{\ddd}{\nab}

\def\c{\cdot}

\def\Da{{^{(A)}\hskip-.15 pc \D}}

\newcommand{\distb}{d_\backmg}
\newcommand{\hmcK}{\widehat \mcK}
\newcommand{\ptcx}[1]{\ptc{#1}}

\def\hot{\widehat{\otimes}}
\def\lot{\mbox{l.o.t.}}

\renewcommand{\div}{\mbox{div }}
\newcommand{\divv}{\mbox{div} }
\def\err{\mbox{Err}}
\newcommand{\lapp}{\mbox{$\bigtriangleup  \mkern-13mu / \,$}}

\newcommand{\piX}{\,^{(X)}\pi}
\def\diag{{\mbox{diag}}}
\def\Flux{\mbox{Flux}}
\def\En{\mbox{En}}
\def\ub{{\underline u}}
\def\dual{{\,^\star \mkern-3mu}}
\def\2{{\overline 2}}
\def\NI{\noindent}
\def\Cb{\underline C}

\newcommand{\beqa}{\begin{eqnarray}}
\newcommand{\eeqa}{\end{eqnarray}}

There are two points which have not been handled properly in
the current work, published as~\cite{Chstatic}.
\renewcommand{\changedX}{X}%
\renewcommand{\ker}{\Ker}%
\renewcommand{\mcM}{M}%
\renewcommand{\hyp}{\Sigma}%
\renewcommand{\fourg}{\mathbf{g}}%
\renewcommand{\ohyp}{\bar \Sigma}%
\renewcommand{\hypot}{ \mathring{  \hyp}_t}%
\renewcommand{\hypoz}{ \mathring{\hyp}_0}%
\renewcommand{\ohypoz}{\overline{\hypoz}}%
\renewcommand{\ohypot}{\overline{\hypot}}%
\renewcommand{\hypo}{\mathring \hyp}%
\renewcommand{\ohypo}{\overline{\hypo}}%
First, it has been pointed out to me by Jo\~ao~Lopes~Costa that
neither the original proof, nor that given in~\cite{Chstatic},
of the Vishveshwara-Carter Lemma, takes properly into account
the possibility that the hypersurface $\mcN$ of
\cite[Lemma~4.1]{Chstatic} could fail to be embedded when it is
degenerate. This problem arises whether or not the horizon is
degenerate, since we do not know a priori whether or not $\mcN$
has anything to do with the horizon. This issue is taken care
of by~\cite{ChGstaticity} under the assumption of global
hyperbolicity of the domain of outer communications. I am
grateful to Jo\~ao for pointing out the problem, and for useful
remarks on previous versions of this \emph{corrigendum}.

A (wrong) solution to this problem has been proposed in   the
arXiv version 2 of~\cite{Chstaticelvacarxiv} (that paper was
intended as arXiv version 2 of~\cite{Chstatic}, but has been
posted as version 2 of~\cite{Chstaticelvacarxiv} by an error of
manipulation). The idea was to show that the family of
hypersurfaces covering the set where the static Killing vector
becomes null contains an outermost closed and embedded
hypersurface. A family of curves in a plane that does not
contain such a curve is drawn in Figure~\ref{Fnogo}: the only
reasonable candidate for an outermost curve there is the curve
of infinite length, which is embedded, but does not form a
closed subset of the plane. The example shows that the
strategy proposed in the addendum to~\cite{Chstaticelvacarxiv} has no chance of succeeding.%
\begin{figure}[h]
\begin{center} {
\resizebox{2in}{!}{\includegraphics{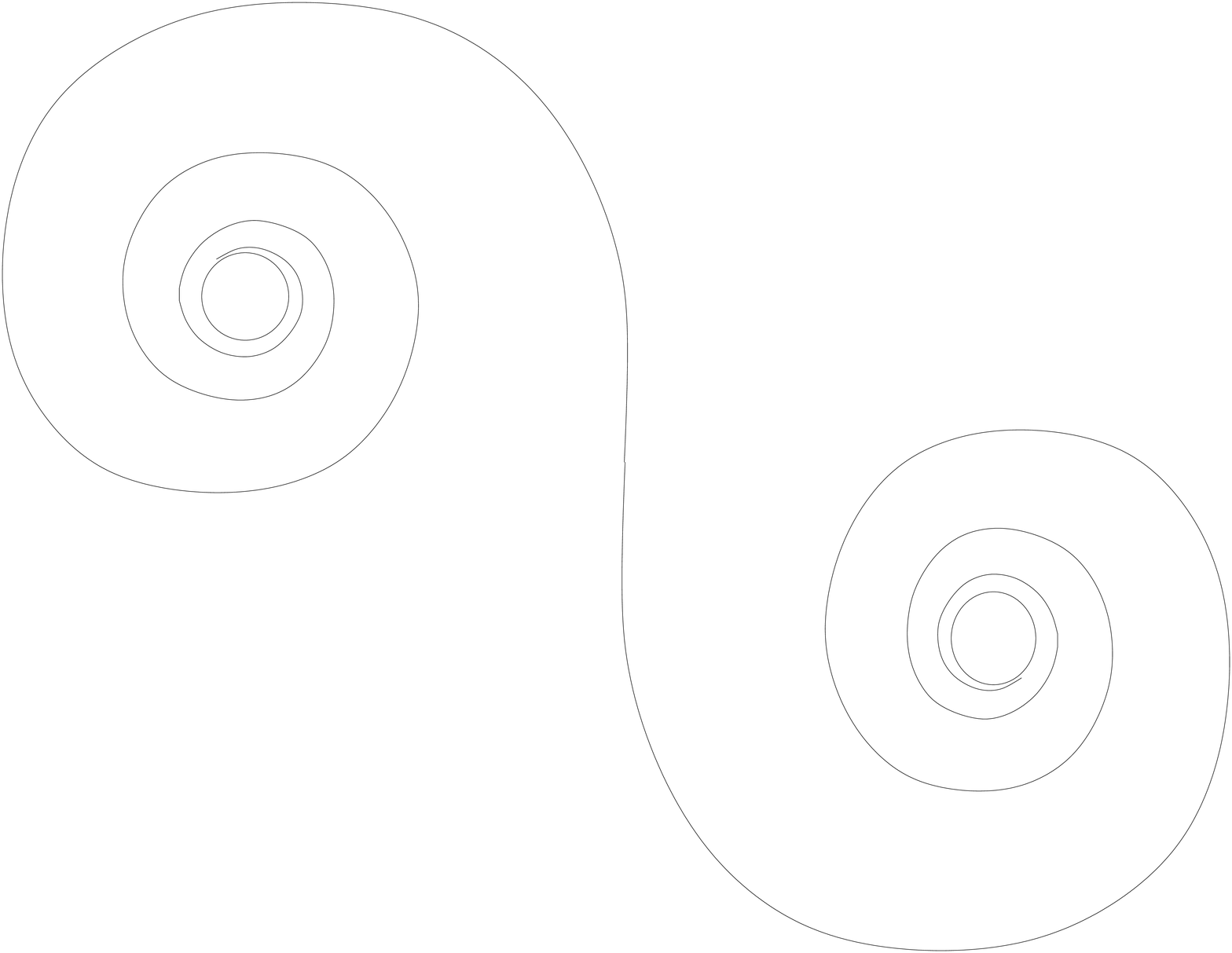}}
}
\caption{A family of three embedded curves in the plane consisting of two circles, together with a
curve of infinite length that spirals towards the circles.
\label{Fnogo}}
\end{center}
\end{figure}

We note a corrected version of the Vishveshwara--Carter
Lemma~\cite{Vishveshwara,CarterJMP}:

\newcommand{\hmcN}{\,\,\widehat{\!\!\mcN}}
\begin{Lemma}
  \label{lcnewx}
Let $(\mcM,\fourg)$ be a smooth space--time with complete,
static Killing vector $X$, set
\bel{Wdef} W:=-\fourg_{\alpha\beta}X^\alpha
  X^\beta
 \;.
\ee
Then

\begin{enumerate}
\item  $\{W=0\}\cap \{X\ne 0\}$ is a union of integral
    leaves of the distribution $X^\perp$, which are totally
    geodesic within $\mcM\setminus \{X=0\}$.
 \item \label{lcnewxP2} Each connected component of
 $$
 \{W=0\}\cap \{dW\ne 0\}\cap \{X\ne 0\}
 $$
 is a smooth, embedded, locally totally geodesic null
 hypersurface $\mcN$, with the Killing vector field $X$
 normal to ${\mcN}$.
\end{enumerate}%
\end{Lemma}

\begin{Remark}
{ \rm As the surface gravity is  constant in electro-vacuum,
point 2 covers adequately non-degenerate vacuum or
electro-vacuum Killing horizons, but does not apply to
degenerate ones. In~\cite[Lemma~4.1]{Chstatic}, the conclusions
of point \ref{lcnewxP2}. were claimed for each connected
component of the set
$$
\{W=0\}\cap \{X\ne 0\} \cap \partial \{W<0\}
\;,
$$
but the justification given there does not seem to be
sufficient, and we do not know whether or not the result is
correct without further hypotheses.
 }
\end{Remark}

{\noindent \sc Proof:} The staticity condition $X^\flat \wedge
dX^\flat=0$ and  the Frobenius theorem~\cite[Section
9.1]{Hicks} show that
$$
 \mcO:=M\setminus\{X=0\}
$$
is foliated by immersed, but not necessarily embedded,
hypersurfaces which are normal to $X$. Let
$$\hat S_p\subset \mcO
$$
denote the maximally extended integral leaf of the distribution
$X^\perp$ passing through $p$.

Let $p\in \{W=0\}\cap \{X\ne 0\} $ and  let $\gamma$ be an
affinely parameterised geodesic starting at $p$ with initial
tangent $\dot \gamma(0)$ normal to $X$. A standard calculation
along $\gamma$ shows that
$$
\frac {d (\fourg(\dot \gamma,X))}{ds}= \dot \gamma^\mu \nabla_\mu(\dot \gamma^\nu X_\nu)
 =\underbrace{\dot \gamma^\mu \nabla_\mu \dot \gamma^\nu}_{=0} X_\nu+
  \dot \gamma^\nu \dot \gamma^\mu \nabla_\mu X_\nu =
  \dot \gamma^\nu \dot \gamma^\mu\underbrace{\nabla_{(\mu} X_{\nu)}}_{=0} =
 0
 \;,
$$
hence $\dot \gamma$ remains normal to $X$, implying that
$\hS_p$ is locally totally geodesic, as claimed. Clearly
$\gamma$ can   exit $\hS_p$ only where that  leaf ceases to be
defined, namely at zeros of $X$. Hence the $\hS_p$'s are
totally geodesic within $\mcM\setminus \{X=0\}$.

The staticity condition  can be rewritten as
$$
2 \nabla_{[\mu} X^\alpha X_{\nu]} =  X^\alpha \nabla_{[\mu} X_{\nu]}\
,
$$
which, after a contraction with $X_\alpha$, gives
\begin{equation}
  \label{eq:propx}
   \nabla_{[\mu}W X_{\nu]} = W \nabla_{[\mu} X_{\nu]}\
, \qquad W\equiv X_\alpha X^\alpha\ .
\end{equation}%
Let $\ell_\mu$ be any smooth covector field on $\mcO$ such that
$\ell_\mu X^\mu=1$ and let $\gamma$ be any differentiable curve
contained in a leaf $\hS_p$ such that $W(\gamma(0))=0$;
contraction of \eq{eq:propx} with  $\dot \gamma^\mu \ell ^\nu$
gives
\begin{equation}
  \label{eq:propx2}
  \frac{dW}{ds} = \dot \gamma^\mu \nabla _\mu W = 2 W \dot \gamma^\mu \ell ^\nu \nabla_{[\mu} X_{\nu]}
\ .
\end{equation}
Uniqueness of solutions of ODEs implies that $W\circ \gamma=0$,
and we conclude that $\hS_p\subset \{W=0\} $. This shows that
if $\hS_q\cap \{W=0\}\ne\emptyset$ then $W\equiv 0$ on $\hS_q$.
Hence $\{W=0\}\setminus \{X=0\}$ is a union of leaves of the
$\hS_p$ foliation.

At those points at which $d W$ does not vanish, the set
$\{W=0\}$ is smooth, embedded hypersurface, and the proof is
complete.
\qed

Next,  the degenerate case in  Boyer's theorem~\cite{Boyer} has
been quoted incorrectly in~\cite[Theorem~3.1]{Chstatic}: In
spite of what is said there, there exist Killing vectors which
have zeros at the closure of a degenerate horizon. The
Minkowskian Killing vector
\bel{Xbad}
 X = t\partial_x + x \partial_t + x\partial_y - y \partial _x = (t-y) \partial_x + x(\partial_t+\partial_y)
\ee
illustrates well the problem at hand. $X$ vanishes at
$\{t=y\;,\ x=0\}$, and is null on $\{t=y\;,\ x\ne 0\}$. Recall
that a Killing horizon associated to $X$ is a null hypersurface
$\mcN$ on which $X$ is null,  tangent to $\mcN$. So, in this
case, $\mcN$ has two connected components
$$
 \mcN^\pm:= \{t=y\;, \pm x>0\}
 \;.
$$
A key for the proof of~\cite[Theorem~1.1]{Chstatic} is
Proposition~3.2 there, which is \emph{wrong} without the
supplementary hypothesis that the Killing vector $X$ has no
zeros on $\partial \bar\Sigma $. Indeed, let $\hyp=\{t=0,y>
0\}$ in Minkowski space-time $(\R^4,g)$, and let $X$ be given
by \eq{Xbad}. To
see that $\hyp$ equipped with the ``orbit space metric"%
\footnote{As already emphasised in~\cite{Chstatic}, the metric
$h$ should \emph{not} be thought of as the ``metric on the
space of orbits", as we are not assuming anything about the
manifold character of this last space; similarly transversality
of   $\chX$ to $\hyp$ is \emph{not} assumed.}
$$
\forall Z_1,Z_2 \in T\hyp\qquad
h(Z_1,Z_2) = g(Z_1,Z_2) - \frac{g(X,Z_1)g(X,Z_2)}{g(X,X)}
$$
contains finite-proper length spacelike curves which reach the
boundary $\partial \bar \Sigma$, consider the curve
$$
 (0,\infty)\ni s \mapsto \gamma(s)=(t=x=z=0\;, \ y=s)\in \hyp
 \;.
$$
Then $g(X,\dot \gamma)=0$, so $h(\dot \gamma,\dot \gamma) =
g(\dot \gamma, \dot \gamma) =1$, and the boundary $y=0$ lies at
$h$--distance along $\gamma$ equal to $s$ from any point
$\gamma(s)$ on $\gamma$. But then this boundary lies to finite
$h$--distance from any point $p \in \hyp$, as one can reach
$\partial\bar\hyp$ from $p$ by first going to $\gamma$, and
then following $\gamma$ until $\partial \overline \hyp $ is
reached. Here $\partial \bar \hyp $ is not compact, but this
seems irrelevant for the issue at hand.

More significantly, in this example $X$ is spacelike near and
away from $\{t=z\}$, so that   $h$ is \emph{Lorentzian} there,
while the orbit space metric \emph{is} Riemannian in the
context of the analysis of~\cite{Chstatic}. This observation is
the key to the proof below that such zeros do not exist on
boundaries $\partial\{\fourg(X,X)=0\}$ in static space-times.

\medskip
\renewcommand{\changedX}{X}%

Let us thus show nonexistence of the offending zeros%
\footnote{Zeros of $X$ occurring at non-degenerate components
of $\partial \bar \Sigma$ are allowed in
\cite[Theorem~1.1]{Chstatic}.}
of $X$ under the hypotheses of~\cite[Theorem~1.1]{Chstatic}.
Recall that it is assumed there that a vacuum space-time
$(\mcM,\fourg)$ has a hypersurface-orthogonal Killing vector
$\changedX$ which is timelike on a spacelike hypersurface
$\Sigma$, and vanishes on its boundary $\partial \bar \hyp=\bar
\hyp \setminus \hyp$, which is assumed to be a compact
two-dimensional topological manifold. Now, as shown
in~\cite{ChCo}, the set, say $\mcE$, where
$\fourg(\changedX,\changedX)$ vanishes, is foliated by locally
totally geodesic null hypersurfaces, away from the points where
$\changedX$ vanishes. Hence each leaf of $\mcE$ is smooth on an
open dense set, so $\partial\hyp$ is smooth on the open dense
subset of $\partial \hyp$ consisting of  points at which
$\changedX$ does not vanish. Note that $\mcE$ might fail to be
embedded in general, but this is irrelevant for the proof here
because $\partial \hyp$ is a compact embedded topological
manifold by hypothesis. In vacuum, on every smooth leaf of
$\mcE$, and hence on every smooth component of $\partial \hyp$,
the surface gravity $\kappa$ is constant (see, e.g.,
\cite[Theorem~2.1]{RaczWald2}). It follows that the problem
with the incorrect~\cite[Theorem~3.1]{Chstatic} is avoided by
the following result:

\begin{Proposition}\label{Ponce2x}
Let $\changedX$ be a  Killing vector field,  and suppose that
\bel{boundeqOm}
 \Omega:=\partial\{p\in \mcM \ | \
\fourg(\changedX,\changedX)<0\}
 \;.
  \ee
is a topological hypersurface. Suppose that
\begin{enumerate}
 \item either $\changedX$ is hypersurface-orthogonal and
     $\Omega$ has vanishing surface gravity wherever
     defined,
 \item or $\Omega$ is differentiable.
\end{enumerate}
 Then $\changedX$
has no zeros on ${\Omega}$.
\end{Proposition}

\proof
The proof here is an adaptation to space-dimension $n=3$ of a
similar result proved in all dimensions in~\cite{ChHighDim}.
Let $\changedX$ be a non-trivial Killing vector, and suppose
that $\changedX$ vanishes at $p\in \Omega$. Consider the
anti-symmetric tensor $\lambda_{\mu\nu}=\nabla_\mu
\changedX_\nu|_p$; from~\cite[Section~7.2]{Hall:book} or from
\cite{Boyer} we have the following alternative:
 \begin{enumerate}

\item \label{P2} There exists at $p$ an orthonormal frame
    $e_c$, $c=0,\ldots,3$, with $e_0$ timelike, such that
in this frame we have
 \bel{can2}
 \lambda_{cd}=\left(\begin{tabular}{cccc}
               0 & $a $& 0 & 0 \cr $-a$ & 0 & $a$ & 0 \cr 0
                    & $-a$ &0& $0$ \cr 0 & 0 & $0$ & 0 \cr
              \end{tabular}
              \right) \;, \ee
with $a \ne 0$ unless $\changedX\equiv 0$.  Let $\mcU$ be a
geodesically convex neighborhood of $p$ covered by normal
coordinates $(t,x,y,z)$ centred at $p$, and associated to
$e_a$. Because the flow of $X$ maps null geodesics to null
geodesics, we have
\bel{Ycan2} \changedX= a\Big(x\partial_t+(t-y)\partial_x
 +x\partial_y \Big) \;. \ee
This, together with elementary properties of normal
coordinates, implies
\bel{Ynorm2} \fourg(\changedX,\changedX) = a^2(t-y)^2+
 O((t^2+x^2+y^2+z^2)^2) \;.
\ee
It follows from \eq{Ycan2} that $\changedX$ is tangent to
the two hypersurfaces
$$
 \mcN^\pm=\{t=y\;,\ \pm x>0\}
 \;,
$$
non-vanishing there.

Assume that $\changedX$ is hypersurface-orthogonal.
Consider any point $q\in \Omega$ at which $\changedX$ does
not vanish. By Lemma~\ref{lcnewx} the hypersurface $\Omega$
is smooth near $q$, and any geodesic $\gamma$ initially
normal to $\changedX_q$ stays on $\Omega$, except perhaps
when it reaches a point at which $\changedX$ vanishes.

So, suppose that $\gamma$ is such a geodesic from $q\in
\Omega$ to $p$, with $p$ being the first point on $\gamma$
at which $\changedX$ vanishes. If $\dot t\ne \dot y$ at
$p$, \eq{Ynorm2} shows that $\changedX$ is spacelike  along
$\gamma$ near and away from $p$, contradicting the fact
that $\changedX$ is null on $\Omega$. We conclude that
$\dot \gamma$ is tangent at $p$ to the hypersurface
$\{t=y\}$, but then $\gamma\cap \mcU$is included in
$\{t=y\}$. Consequently
\bel{Ominc} \Omega\cap \mcU \subset \{t=y\} \;. \ee
Since $\Omega$ is a topological hypersurface by hypothesis,
we obtain that
\bel{Omesubs} \Omega \cap \mcU = \{t=y\} \;. \ee
(In particular $\Omega$ is smooth near $p$.)

In the case where $\changedX$ is not necessarily
hypersurface orthogonal, but we assume a priori that
$\Omega$ is differentiable, the argument is somewhat
similar, with a weaker conclusion: Let $\gamma\subset
\Omega$ be any differentiable curve, then we must have
$\dot t=\dot y$ at $p$. Since $\Omega$ is a hypersurface,
this implies that
\bel{Omesubs2} T_p\Omega=T_p\{t=y\} \;. \ee
So, while \eq{Omesubs} does not necessarily hold, the
tangent spaces coincide at $p$ in both cases.

Consider, now any differentiable curve $\sigma$ through $p$
on which $\dot t\ne \dot y\ne 0$ at $p$. As already noted,
\Eq{Ynorm2} shows that on $\sigma$ the Killing vector
$\changedX$ is spacelike near and away from $p$. By
\eq{Omesubs2} such curves are transverse to $\Omega$, which
shows  that there exist points arbitrarily close to
$\Omega$ at which $\changedX$ is \emph{spacelike} on both
sides of $\Omega$. This contradicts \eq{boundeqOm}, and
shows that this case cannot happen under our hypotheses.

Note that $\changedX$ is \emph{null future directed} on
$\mcN^+$, \emph{null past directed} on $\mcN^-$,%
\footnote{This fact can be used to given an alternative
justification that $\changedX$ has no zeros on degenerate
components of $\pdoc$ if $\doc$ is chronological, using the
fact that Killing orbits through $\doc$ are then
future-oriented in the sense of~\cite{ChCo}. But the
current argument does not need the chronology hypothesis.}
and vanishes on the set
$$
 \mcY:= \{x=0\;, \ t=y\}= \overline{\mcN^+} \cap \overline{\mcN^-}  \;.
$$
 \item There exists at $p$ an orthonormal frame $e_c$,
     $c=0,\ldots,3$, with $e_0$ timelike, such that in this
     frame we have
 \bel{can}
 \lambda_{cd}=\left(\begin{tabular}{cccc}
               0 & $a $& 0 & 0 \cr $-a$ & 0 & 0 & 0 \cr 0 &
                    0 & 0 & $b$ \cr 0 & 0 & $-b$ & 0 \cr
              \end{tabular}
              \right) \;, \ee
with $a^2+b^2\ne 0$ unless $\changedX\equiv 0$. As before,
in normal coordinates $(t,x,y,z)$ centred at $p$, and
associated to $e_a$, we then have
\bel{Ycan1} \changedX= a(t\partial_x+x\partial_t)+b
(y\partial_z-z\partial_y)
 \;, \ee
leading to
\bel{Ynorm} \fourg(\changedX,\changedX) = a^2 (t^2-x^2)+
 b^2(y^2+z^2) + O((t^2+x^2+y^2+z^2)^2) \;.
\ee
Suppose, first, that $a=0$. Then $\Ker
\lambda=\Span\{\partial_t,\partial_x\}|_p$. Now, because
the flow of a Killing vector maps geodesics to geodesics,
$\changedX$ vanishes on every geodesic $\gamma$ with
$\gamma(0)=p$ such that $\dot \gamma (0)\in \Ker \lambda$.
So $\changedX$ vanishes throughout the timelike
hypersurface $\{y=z=0\}$. At every point $q$ of this
hypersurface, in adapted normal coordinates centred a $q$
the tensor $\nabla_c \changedX_d|_q$ takes the form
\eq{can} with $a=0$. This implies that $\changedX$ is
spacelike or vanishing throughout a neighborhood of $p$, so
$a=0$ cannot occur.

  If $\Omega$ is differentiable
at $p$, an argument very similar to the one above shows
that
$$
 T_p\Omega \subset E_+\cup E_-\;, \quad \mbox{where} \ E_\pm:=\{\dot t=\pm \dot x\}
 \;.
$$
So either $T_p\Omega=E_+$ or $T_p\Omega=E_-$. But, the
curves with $\dot t=  2\dot x$ at $p$  are transverse both
to $E_-$ and to $E_+$, with $\changedX$  spacelike on those
curves near and away from $p$ on both sides of $E_\pm$,
contradicting the definition of $\Omega$. Assuming
differentiability of $\Omega$ we are done.

We continue with an analysis of the static case, and claim
that $ab\ne 0$ is not possible. Indeed, let $X^\flat$ be
the field of one-forms defined as
$X^\flat=\fourg(X,\cdot)$. Then
\beal{Xflat1}
X^\flat & = &   a(t dx-xdt) + b(ydz-zdy)+
 O((t^2+x^2+y^2+z^2)^{3/2}) \;,\phantom{xxx}
 \\
 dX^\flat & = &   2a \, dt\wedge   dx + 2  b\, dy\wedge dz
 + O( t^2+x^2+y^2+z^2 ) \;,
\eea
and the staticity condition $X^\flat \wedge dX^\flat =0$
gives $ab=0$.

It remains to consider $b=0$.
Arguments similar to the ones already given show that
$$
 \Omega\cap\mcU \cap \{y=z=0\;, \ t=\pm x\}\ne \emptyset
 \;. $$
In this case from \eq{Ynorm} we have
$$
 d\left(\fourg(\changedX,\changedX)\right) =2a^2(t dx-xdt) +
 O((t^2+x^2+y^2+z^2)^{3/2}) \;.
$$
 Comparing with \eq{Xflat1} with $b=0$ at points lying on
the surface $\{y=z=0\;, \ t=\pm x\}$, with $|x|$
sufficiently small, we conclude that this case cannot occur
if $\Omega$ is degenerate, and the proof is complete.
\qed
\end{enumerate}

\bibliographystyle{amsplain}
\bibliography{../references/hip_bib,%
../references/reffile,%
../references/newbiblio,%
../references/bibl,%
../references/howard,%
../references/bartnik,%
../references/myGR,%
../references/newbib,%
../references/Energy,%
../references/netbiblio}

\end{document}